\documentclass[twocolumn]{aastex63}
\usepackage{apjfonts}
\usepackage{gensymb}

\received{2020 September 15}
\revised{2020 November 11}
\accepted{2020 December 2}
\submitjournal{ApJ}

\shorttitle{Difference in chemical composition between the two RCs}
\shortauthors{Lim et al.}

\begin{document}

\title{
Difference in chemical composition between the bright and faint red clump stars in the Milky Way bulge  
}

\correspondingauthor{Dongwook Lim} 
\email{dongwook.lim@uni-heidelberg.de}

\author[0000-0001-7277-7175]{Dongwook Lim}
\affiliation{Zentrum f\"ur Astronomie der Universit\"at Heidelberg, Astronomisches Rechen-Institut, M\"onchhofstr. 12-14, 69120 Heidelberg, Germany}
\affiliation{Center for Galaxy Evolution Research \& Department of Astronomy, Yonsei University, Seoul 03722, Republic of Korea}

\author{Young-Wook Lee}
\email{ywlee2@yonsei.ac.kr}
\affiliation{Center for Galaxy Evolution Research \& Department of Astronomy, Yonsei University, Seoul 03722, Republic of Korea}

\author[0000-0002-9859-4956]{Andreas Koch}
\affiliation{Zentrum f\"ur Astronomie der Universit\"at Heidelberg, Astronomisches Rechen-Institut, M\"onchhofstr. 12-14, 69120 Heidelberg, Germany}

\author{Seungsoo Hong}
\affiliation{Center for Galaxy Evolution Research \& Department of Astronomy, Yonsei University, Seoul 03722, Republic of Korea}

\author{Christian I. Johnson} 
\affiliation{Space Telescope Science Institute, 3700 San Martin Drive, Baltimore, MD 21218, USA}

\author[0000-0002-0432-6847]{Jenny J. Kim}
\affiliation{Zentrum f\"ur Astronomie der Universit\"at Heidelberg, Astronomisches Rechen-Institut, M\"onchhofstr. 12-14, 69120 Heidelberg, Germany}

\author{Chul Chung}
\affiliation{Center for Galaxy Evolution Research \& Department of Astronomy, Yonsei University, Seoul 03722, Republic of Korea}

\author{Mario Mateo}
\affiliation{Department of Astronomy, University of Michigan, Ann Arbor, MI 48109, USA} 

\author{John I. Bailey, III}
\affiliation{Department of Physics, UCSB, Santa Barbara, CA 93016, USA}

\begin{abstract}
The double red clump (RC) observed in color-magnitude diagrams of the Milky Way bulge is at the heart of the current debate on the structure and formation origin of the bulge. This feature can be explained by the difference between the two RCs either in distance (``X-shaped scenario'') or in chemical composition (``multiple-population scenario''). Here we report our high-resolution spectroscopy for the RC and red giant branch stars in a high-latitude field ($b \sim -8.5\degree$) of the bulge. We find a difference in [Fe/H] between the stars in the bright and faint RC regimes, in the sense that the bright stars are enhanced in [Fe/H] with respect to the faint stars by 0.149 $\pm$ 0.036 dex. The stars on the bright RC are also enhanced in [Na/Fe] but appear to be depleted in [Al/Fe] and [O/Fe], although more observations are required to confirm the significance of these differences. Interestingly, these chemical patterns are similar to those observed among multiple stellar populations in the metal-rich bulge globular cluster Terzan~5. In addition, we find a number of Na-rich stars, which would corroborate the presence of multiple populations in the bulge. Our results support an origin of the double RC from dissolved globular clusters that harbor multiple stellar populations. Thus, our study suggests that a substantial fraction of the outer bulge stars would have originated from the assembly of such stellar systems in the early phase of the Milky Way formation.
\end{abstract}



\section{Introduction} \label{sec:intro}
The Milky Way (MW) is the best laboratory for studying the formation and evolution of a galaxy and its individual components, such as the disks, halo, and bulge. Recently, an increasing number of photometric and spectroscopic surveys of resolved stars allows for a better understanding of the assembly history of the MW. For the bulge component, several dedicated surveys were carried out or either are in progress or in preparation, each with the aim of revealing its origin and structure. For instance, the Bulge Radial Velocity Assay (BRAVA; \citealt{Howard2008,Kunder2012}), the Abundances and Radial velocity Galactic Origins Survey (ARGOS; \citealt{Freeman2013}), and the GIRAFFE Inner Bulge Survey (GIBS; \citealt{Zoccali2014, Zoccali2017}) dissected the complex chemo-dynamical populations among bulge stars. The Apache Point Observatory Galactic Evolution Experiment (APOGEE; \citealt{Majewski2017, Zasowski2019}) presents detailed chemical abundances for a large number of stars in the bulge. Nonetheless, compared to the halo and the disks, our understanding of the MW bulge is less clear because of the higher and differential extinction and severe contamination of other stellar components in the Galactic plane. In particular, it is still debated whether the MW bulge is formed from the hierarchical merging, buckling instability of the bar/disk or other mechanisms \citep[see, e.g.,][]{Nataf2017}. The hierarchical merging during the early phase of galaxy formation would produce `classical bulge', whereas `pseudo bulge' is formed through secular evolution of the bar/disk. These two models have very different properties on the structure and kinematics of the bulge, where the `classical bulge' and `pseudo bulge' show a spheroidal-like and boxy/peanut-like shapes, respectively \citep[see][]{Athanassoula2005, Barbuy2018}. It is reported that the MW bulge has characteristics of both types of bulge based on various sources, including RR Lyrae stars and Miras \citep{Kunder2020,Grady2020}. However, the question on the contribution of each bulge component is still open.

In this regard, the double red clump (RC) feature discovered in high-latitude fields of the Galactic bulge plays a crucial role in the study of its structure and formation \citep{McWilliam2010, Nataf2010}. Since RC stars are thought to have a narrow luminosity distribution, they have been widely used as a distance indicator \citep{Paczynski1998}. The presence of a double RC, therefore, was initially suggested as evidence for a giant X-shaped structure with the assumption that the bright RC (bRC) and the faint RC (fRC) have different distance from the Sun \citep{McWilliam2010, Li2012, Wegg2013}. In the `pseudo bulge' model, such an X-shaped structure can be formed from the buckling instability of the bar \citep{Patsis2002, Bureau2006}. 

\citet{Lee2015}, however, casted doubt on the same intrinsic luminosity of the bright and the faint RC stars. They reproduced the double RC feature within a scenario, where the two RC features emerge from two stellar populations with different chemical compositions (including He-abundance). The presence of multiple populations is now a well-established feature that is notably found in globular clusters (GCs) and pertains to their formation mechanisms \citep[e.g.][]{Piotto2015,Bastian2018}. In general, the second and later-generation (G2\textsuperscript{+}) stars in a GC are enhanced in He and certain light elements (N, Na, Al; and Fe in some cases) compared to the first-generation (G1) stars, which are, in turn, depleted in those elements while showing enhancements of O and Mg \citep[see, e.g.,][and references therein]{Bastian2018}. It is worth noticing that the ensuing element bimodalities and (anti-)correlations are intrinsic to the GCs and not found in any other stellar population \citep{Villanova2017}. The double RC observed in color-magnitude diagrams (CMDs) of the MW bulge can thus also be explained by the chemical composition of G1 star making up the fRC, and G2\textsuperscript{+} populating the bRC (see Figure~1 of \citealt{Lee2019}). This ``multiple-population scenario'', therefore, further implies that a significant fraction of outer bulge stars were originally formed in proto-GC or GC-like environments and then dissolved in the bulge (see also \citealt{Schiavon2017b}).

The multiple-population scenario conforms with a hierarchical merging or accretion origin of the outer bulge \citep[see][]{Lee2019}, while the X-shape scenario rather supports a `pseudo bulge'. Therefore, various studies have been conducted to reveal its origin, using chemical evolution models and distance information \citep[e.g.,][]{Joo2017, Lopez-Corredoira2019}. An early study by \citet{DePropris2011} did not find any discernible difference between the two RCs, neither in kinematics, nor in their abundances derived from low-resolution Mg$b$-index measurements. This was contrasted by \citet{Lee2018}, who found a clear difference in CN-band strength between stars in the bright and faint RC regimes through low-resolution spectroscopy. Since the CN-band is an effective tracer of second-generation GC stars (\citealt{Lim2015, Lim2017,Koch2019}), this finding strengthens the multiple population origin of the double RC. \citet{Lee2019} also revealed that bright red giant branch (RGB) stars in the outer bulge can be divided into two groups according to Na abundance, supporting the presence of multiple stellar populations that likely originated from proto-GCs. 

Here we investigate the detailed chemical composition of RC stars in the MW bulge through  high-resolution spectroscopy, not only to investigate the presence of multiple stellar populations amongst the two RCs, but also to examine their origin. This paper is organized as follows. In Section~\ref{sec:obs}, we describe the observation and data reduction process. The abundance measurements are laid out in Section~\ref{sec:abun}. We compare the metallicity of stars in the two RCs in Section~\ref{sec:metal}, while results for other elements are presented in Section~\ref{sec:other}. We also compare our results with the previous study for CN and CH bands in Section~\ref{sec:lrs}. Finally, possible scenarios for the double RC are discussed in Section~\ref{sec:dis}.


\section{Observation and data reduction} \label{sec:obs}
We have observed RC and RGB stars in the high-latitude bulge field at ($l \sim -1 \degree$, $b \sim -8.5\degree$), where the double RC is most prominently observed and the contamination of the Galactic bar should be negligible. In our previous study, we measured CN-band strengths of stars in this field through low-resolution spectroscopy for the same purpose \citep[see][]{Lee2018}. Here, we aim at deriving the more detailed chemical composition of stars, including Fe, Na, and Al abundances, from higher resolution spectroscopy (R $\sim$ 21,000). We selected $\sim$ 450 target stars from the 2MASS All-Sky Point Source Catalog \citep{Skrutskie2006}, and many of them ($\sim$ 300) overlap with our previous low-resolution sample for the comparison. 

The high-resolution spectra were obtained using the Michigan/Magellan Fiber System (M2FS; \citealt{Mateo2012}) on the Magellan-Clay 6.5m Telescope at the Las Campanas Observatory. We used the ``Hires'' mode of  M2FS with an order blocking filter ``Bulge$\_$GC1'' \citep{Johnson2015}, which provides wavelength coverage from 6120~{\AA} to  6720~{\AA} at a spectral resolution, $R$ $\sim$ 21,000. The M2FS instrument can host up to 256 fibers in two spectrograph channels covering a field of view of 29.2$\arcmin$. However, we only used 48 fibers for observation, because a single spectrum takes six consecutive orders at our observing configuration. Finally, we designed two metal plates containing all target stars and repeatedly observed the same field with 11 different sets, each of which contains 43 targets. In each observation, we also allocated five fibers on the empty sky in order to obtain a master sky spectrum. The observations were carried out during two observing runs in May 2018 and May--June of 2019. We took at least three 1800-second exposures for each target set in good weather conditions with typical seeing around 0.9$\arcsec$. In addition, 45-second flat-lamp frames and 60-second ThAr lamp frames were taken before or after target exposure for calibration. 

We follow the data reduction procedure of \citet{Johnson2015, Johnson2019}, which employed an identical spectrograph configuration of M2FS to this study. We performed preprocessing of the raw data using the IRAF\footnote{IRAF is distributed by the National Optical Astronomy Observatory, which is operated by the Association of Universities for Research in Astronomy(AURA) under a cooperative agreement with the National Science Foundation.} {\em ccdproc} task for overscan, bias, and dark correction. Then, each spectrum was extracted using the IRAF/HYDRA package via aperture identification, scattered light removal, flat fielding, and wavelength calibration. The final spectrum for each target star was obtained as the median of sky subtracted and continuum normalized spectra of each exposure. In addition, radial velocities (RVs) were measured using the {\em XCSAO} cross-correlation routines within IRAF's RVSAO package \citep{Kurtz1998}. We then estimated signal-to-noise (S/N) ratio as the mean of four measurements at around 6400~{\AA}, 6440~{\AA}, 6650~{\AA} and 6680~{\AA}. The average S/N ratio for all samples is $\sim$ 25 per pixel without a significant trend with magnitude.

\begin{figure}
\centering
\includegraphics[width=0.46\textwidth]{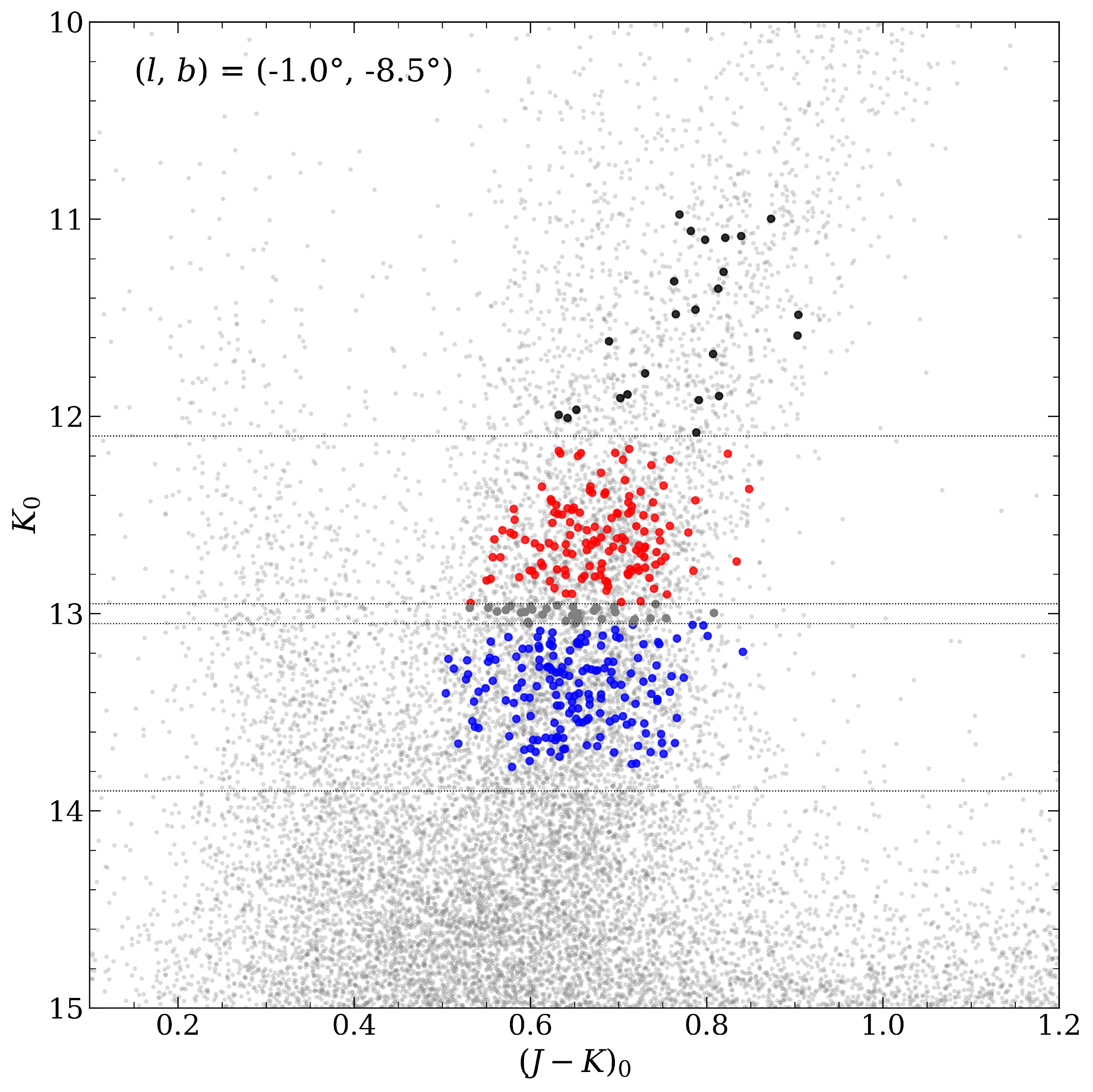}
\figcaption{
 2MASS CMD of stars within our bulge field (0.7$\arcmin$ x 0.6$\arcmin$) in the ($K$, $J-K$) plane. The red, blue, black, and grey circles indicate selected sample stars in bRC, fRC, RGB and twilight zone, respectively (see Section~\ref{sec:metal}).
\label{fig:CMD}
}
\end{figure}

Figure~\ref{fig:CMD} shows the final sample stars (N = 354), excluding bad samples, on the 2MASS CMD.  Note that we derived reddening corrected magnitudes from the dust map of \citet{Schlegel1998} with the extinction laws of \citet{Schlafly2011} and assuming $R$\textsubscript{V} = 3.1.

\section{Abundance analysis} \label{sec:abun}
The chemical abundance analysis was carried out using the 2017 version of the abundance code MOOG \citep{Sneden1973}, which operates under the assumption of Local Thermodynamic Equlilibrium (LTE). First of all, we determined the stellar atmospheric parameters, to wit, effective temperature ($T$\textsubscript{eff}), surface gravity (log~$g$), metallicity ([Fe/H]), and microturbulent velocity ($\xi$\textsubscript{t}), through the usual spectroscopic method \citep[see, e.g.,][]{Koch2016, Johnson2019}. The initial guesses of $T$\textsubscript{eff} and [Fe/H] were obtained using the ATHOS (``A Tool for HOmogenizing Stellar parameters'') code \citep{Hanke2018}, which employs flux ratios for the parameter determination. In the case of log~$g$, we assumed the initial value of 3.5 dex for all sample stars, because our narrow spectral coverage ($\sim$ 600 {\AA}) is insufficient to derive the log~$g$ value with ATHOS, nor does it contain the gravity-sensitive features that ATHOS relies on. Next, we iteratively performed abundance measurements for Fe~\textsubscript{I} and Fe~\textsubscript{II} lines by changing the parameters from the initial guess. The final values were accepted when the trends in the plot of abundance vs. excitation potential and abundance vs. reduced equivalent width (EW), which was estimated from the Gaussian fitting, were removed simultaneously, as well as ionization equilibrium between Fe~\textsubscript{I} and Fe~\textsubscript{II} was satisfied. During the abundance analysis, we applied model atmospheres interpolated from the ATLAS9 grid of Kurucz\footnote{http://kurucz.harvard.edu/grids.html} \citep{Castelli2003}. The atmospheric parameters were firstly estimated using both $\alpha$-enhanced (AODFNEW) and scaled-solar (ODFNEW) opacity distributions, and then we selected final parameters according to the abundance of $\alpha$-elements, namely AODFNEW for stars with [$\alpha$/Fe] $\geq$ 0.15 dex and ODFNEW for stars with [$\alpha$/Fe] $<$ 0.15 dex.  

The abundances of Fe, Si, Ca, Cr, and Ni elements were derived from EWs of each absorption line using the {\em abfind} driver of MOOG. We employed the semi-automatic code developed by \citet{Johnson2014} in order to measure the EWs by fitting single or multiple Gaussian profiles to the lines, respectively. The line list is identical to that used in \citet{Johnson2015}. In addition, we derived abundances of Na, Al, and O at from their transitions at 6154/6160 {\AA} (Na~\textsubscript{I}), 6696/6698 {\AA} (Al~\textsubscript{I}), and 6300 {\AA} ([O~\textsubscript{I}]), respectively, via spectrum synthesis using the {\em synth} driver of MOOG with the recent version of the KURUCZ line list\footnote{http://kurucz.harvard.edu/linelists.html}. We note that spectral synthesis is essential to measure these abundances because these lines are affected by nearby spectral features, such as a Ni-blend and a CN band \citep[see][]{Johnson2019}. In the case of Mg, however, we cannot derive precise abundance due to a broad and strong Ca~\textsubscript{I} autoionization feature in the region of the Mg~\textsubscript{I} absorption lines. The measured atmospheric parameters and abundances are listed in online Table in terms of [X/Fe], calculated assuming the solar abundances of \citet{Asplund2009}.

The measurement uncertainty on the Fe abundance is estimated as the standard error of the mean of the abundances obtained from each absorption lines (N $\sim$ 50). For other elements, we indicate the typical measurement errors in each figure, which are obtained in the same way. Note that the uncertainty on the O abundance is empirically derived from the spectrum synthesis analysis ($\pm$0.10 dex), because this abundance is based only one single line. In addition, in order to examine the systematic abundance errors, which arise from uncertainties in atmospheric parameters, we re-estimate the abundances from eight new atmosphere models with altered stellar parameters ($\Delta$$T$\textsubscript{eff} = $\pm$100$K$, $\Delta$log~$g$ = $\pm$0.15 dex, $\Delta$[Fe/H] = $\pm$0.1 dex, and $\Delta$$\xi$\textsubscript{t} = $\pm$0.1 km s$^{-1}$). The uncertainty of abundance depending on each stellar atmospheric parameter is determined from the difference between the values obtained from the best-fit model and altered model. We perform this examination for five stars with different temperature, from the coolest (4500$K$) to the hottest (5500$K$), and the average errors for each element are listed in Table~\ref{tab:err}.

\begin{deluxetable}{ccccccc}
\tabletypesize{\footnotesize}
\tablecolumns{6}
\tablewidth{0pt}
\tablecaption{Systematic errors due to variation of atmospheric parameters}
\label{tab:err}
\tablehead{
\vspace{-0.4cm} & \colhead{$T$\textsubscript{eff}} & \colhead{log~$g$} & \colhead{[Fe/H]} & \colhead{$\xi$\textsubscript{t}} &	\\
\colhead{Element} \vspace{-0.4cm} & & & & & \colhead{$\sigma$\textsubscript{total}} \\
\vspace{-0.5cm} & \colhead{$\pm$100$K$} & \colhead{$\pm$0.15 dex} & \colhead{$\pm$0.1 dex}	& \colhead{$\pm$0.1 km s$^{-1}$} & \\
}
\startdata
Fe 	& $\pm$0.05 	& $\pm$0.01 	& $\pm$0.01 	& $\mp$0.03 	& 0.06 \\ 
O 	& $\pm$0.03 	& $\pm$0.05 	& $\pm$0.03 	& $\mp$0.02 	& 0.07 \\ 
Na 	& $\pm$0.07 	& $\mp$0.01 	& $<$0.01 	& $\mp$0.02 	& 0.07 \\ 
Al	& $\pm$0.06 	& $<$0.01 	& $<$0.01 	& $\mp$0.02 	& 0.06 \\ 
Si 	& $\mp$0.03 	& $\mp$0.02 	& $\pm$0.02 	& $\mp$0.01 	& 0.04 \\ 
Ca 	& $\pm$0.09 	& $\mp$0.02 	& $<$0.01 	& $\mp$0.04 	& 0.10 \\ 
Cr 	& $\pm$0.13 	& $<$0.01 	& $<$0.01 	& $\mp$0.02 	& 0.04 \\ 
Ni 	& $\pm$0.02 	& $\pm$0.02 	& $\pm$0.02 	& $\mp$0.02 	& 0.04 \\ 
\enddata
\end{deluxetable}


\section{Metallicity difference between the bright and faint RC stars} \label{sec:metal}

\subsection{bright RC vs. faint RC} 
In order to compare the chemical composition between the bright and faint RCs, we divide sample stars into three groups according to their magnitude: a bRC group (12.10 $<$ $K_{0}$ $\leq$ 12.95), a fRC group (13.05 $<$ $K_{0}$ $\leq$ 13.90), and a bright RGB group ($K_{0}$ $\leq$ 12.10), based on the luminosity function of \citet[][see their Figure~1]{Lee2018}. Stars in the `twilight zone' between the two RCs (12.95 $<$ $K_{0}$ $\leq$ 13.05) are excluded from the analysis because this region would be overlapped by stars from both RCs. We note that it is almost impossible to distinguish genuine RC stars from RGB stars in the RC magnitude range, because they have similar properties of color, temperature, and surface gravity \citep[see, e.g.,][]{Hawkins2018}. Amongst a total of 354 stars, 135 and 164 stars belong to the bRC and fRC groups, respectively, and the remaining 24 and 31 stars are in the brighter RGB and twilight zone (see Figure~\ref{fig:CMD}). 

\begin{figure}
\centering
\includegraphics[width=0.45\textwidth]{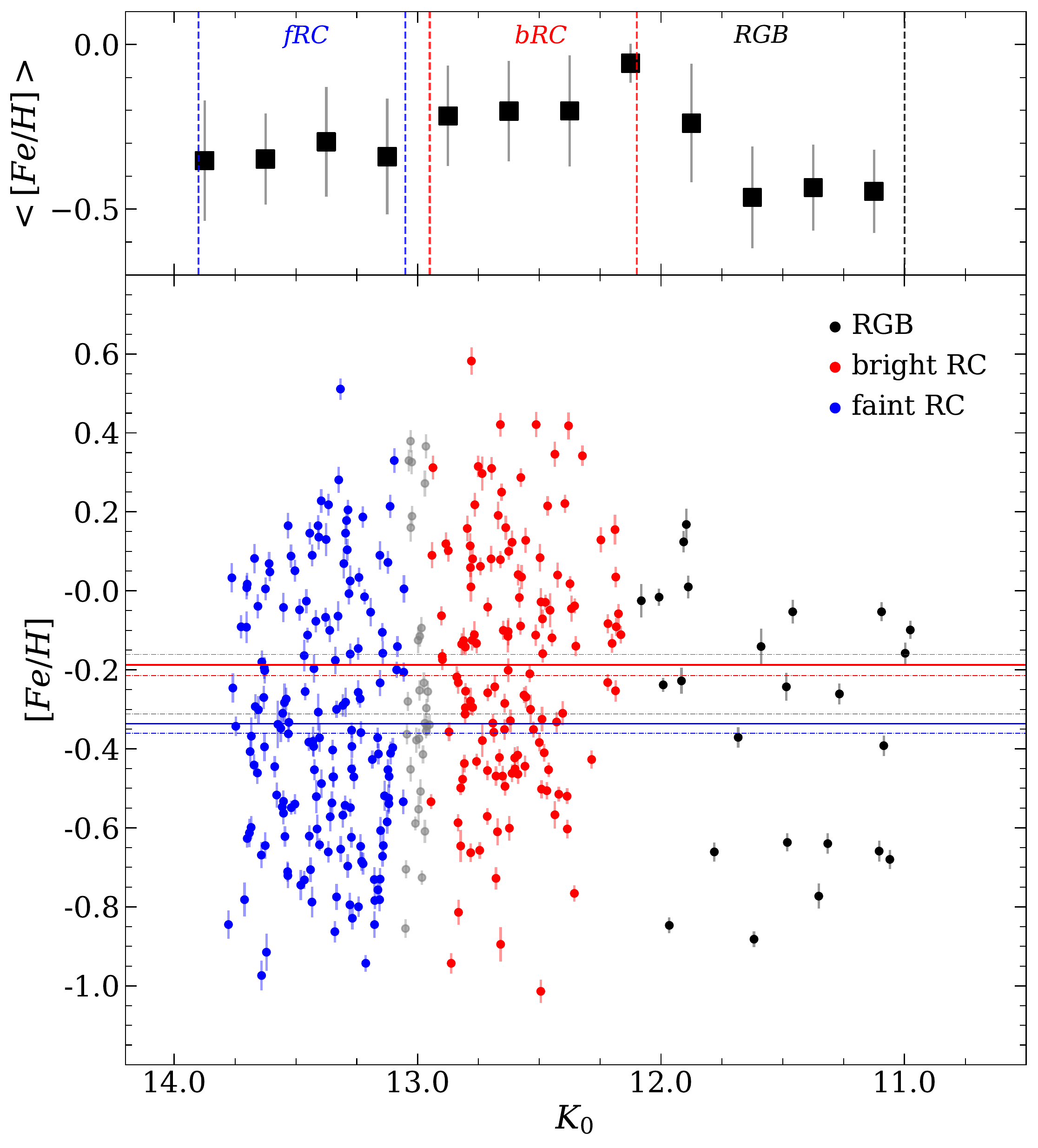}
\figcaption{
Measured [Fe/H] as a function of $K_{0}$ magnitude for our sample stars, where the blue, red, and black circles are stars in the magnitude range of fRC, bRC, and bright RGB, respectively (see Figure~\ref{fig:CMD}). Note that stars in the twilight zone (grey circles) are excluded from the comparison of chemical abundances. The blue and red solid and dotted horizontal lines denote the mean value and the standard error of the mean for each group. The stars belonging to the bRC group are more enhanced in [Fe/H] compared to those in the fRC group ($\Delta$[Fe/H] = 0.149 dex). The upper panel shows the mean [Fe/H] abundance of stars within 0.25 magnitude bins and its standard deviation (vertical bars). The mean values of stars in the bRC regime are slightly higher than those in the fRC regime. The low mean values of RGB stars are unreliable due to the small sample size.
\label{fig:FeH}
}
\end{figure}

The measured [Fe/H] values of stars are shown in Figure~\ref{fig:FeH} as a function of $K_{0}$ magnitude. In this plot, stars belonging to the bRC group (red circles) are more enhanced in [Fe/H] than those in the fRC group (blue circles). The mean [Fe/H] values are $-$0.188 $\pm$ 0.027 dex for the bRC group, and $-$0.336 $\pm$ 0.025 dex for the fRC group, and -0.270 $\pm$ 0.017 dex for the total sample, respectively. The difference between the two RC groups is therefore $\Delta$[Fe/H] = 0.149 $\pm$ 0.036 dex. Although this difference is smaller than the intrinsic dispersion for all samples (0.320~dex), which is estimated following \citet{Piatti2018}, it is four times larger than the error of the mean for the bRC and fRC groups (0.036~dex). Considering the large sample size, the difference in [Fe/H] between the two groups seems reasonable, despite the large metallicity dispersion of the bulge stars. In addition, the upper panel of Figure~\ref{fig:FeH}, which shows the mean and standard deviation of stars divided into twelve bins of 0.25 magnitude, indicates that the stars in the bRC magnitude range have slightly higher mean metallicity than those in the fRC range. A Kolmogorov-Smirnov (KS) test also supports that these two groups are drawn from different distributions with a probability value ($p$-value) of 0.0007 and a KS statistics of 0.23. 

This difference in [Fe/H], however, is not derived only from genuine RC stars, but from all stars in the RC regime, including the background RGB stars in the same regime. Following our previous study, the relative fraction of RC stars to the total stars is estimated to be 34\% and 22\% for the bRC and fRC groups, respectively, from the luminosity function fitting (Figure~1 of \citealt{Lee2018}; see also \citealt{Nataf2013}). This indicates that only $\sim$ 27\% ( = 0.34 $\times$ N\textsubscript{bRC} + 0.22 $\times$ N\textsubscript{fRC} / N\textsubscript{bRC+fRC}) of observed stars in the RC regime would be genuine RC stars. Assuming the identical metallicity distributions for the background RGB stars in the bRC and fRC groups, $\Delta$[Fe/H] of $\sim$ 0.55 dex ( = 0.149 dex $\div$ 27\%) is required between the genuine RC stars in the two groups in order to reproduce the observed difference. We could also confirm this difference from a number of toy models reflecting the number ratio of metal-rich bRC, metal-poor fRC, and background RGB stars. This difference of $\sim$ 0.55 dex is comparable to the intrinsic metallicity variations observed in some peculiar GCs, such as $\omega$-Centauri \citep[{[}Fe/H{]} = $-$1.75 $\sim$ -0.75 dex;][]{Johnson2010}, Terzan~5 \citep[$\Delta${[}Fe/H{]} $\sim$ 0.5 dex;][]{Origlia2011, Massari2014}, and M22 \citep[$\Delta${[}Fe/H{]} $\sim$ 0.3 dex;][]{Da Costa2009, Marino2011}. These GCs are generally suspected as the remnant nuclei of disrupted dwarf galaxies \citep[e.g.,][]{Bekki2003,Da Costa2016}. 

In particular, our bulge field has interesting parallels with a metal-rich GC Terzan~5 ([Fe/H] = $-$0.25 $\sim$ +0.27 dex; \citealt{Origlia2011}). \citet{Ferraro2009} discovered the two distinct RCs on the ($K$, $J-K$) CMD of Terzan~5, which is similar to the double RC feature in the MW bulge. Previous studies have shown that both features are well reproduced within the same multiple population model as invoked here (see Figure~2 of \citealt{Lee2015} and Figure~8 of \citealt{Joo2017}). In addition, the striking similarity in metallicity variation between the bulge and Terzan~5 ($\Delta$[Fe/H] $\sim$ 0.55 dex and $\sim$ 0.52 dex) suggests a association of their formation and evolution mechanism. Thus, our result implies that Terzan~5-like stellar systems could have contributed to the formation of the MW bulge and its double RC feature. We note that Terzan~5 is considered as a very peculiar stellar system\footnote{\citet{Origlia2011,Massari2014} suggested that Terzan~5 is not a true GC because of their complex star formation history and chemical patterns. \citet{Schiavon2017a} also described this system as something in-between GCs and dwarf spheroidal galaxies. There is a same argument for $\omega$-Centauri \citep[e.g.,][]{Bekki2003}. On the other hand, \citet{McKenzie2018} proposed a new scenario that Terzan~5 would be formed by the collision between a metal-poor GC and a giant molecular cloud.} because of its unique properties, such as distinct age and abundance groups \citep[see, e.g.,][and references therein]{Origlia2019}.

\begin{figure*}
\centering
\includegraphics[width=0.85\textwidth]{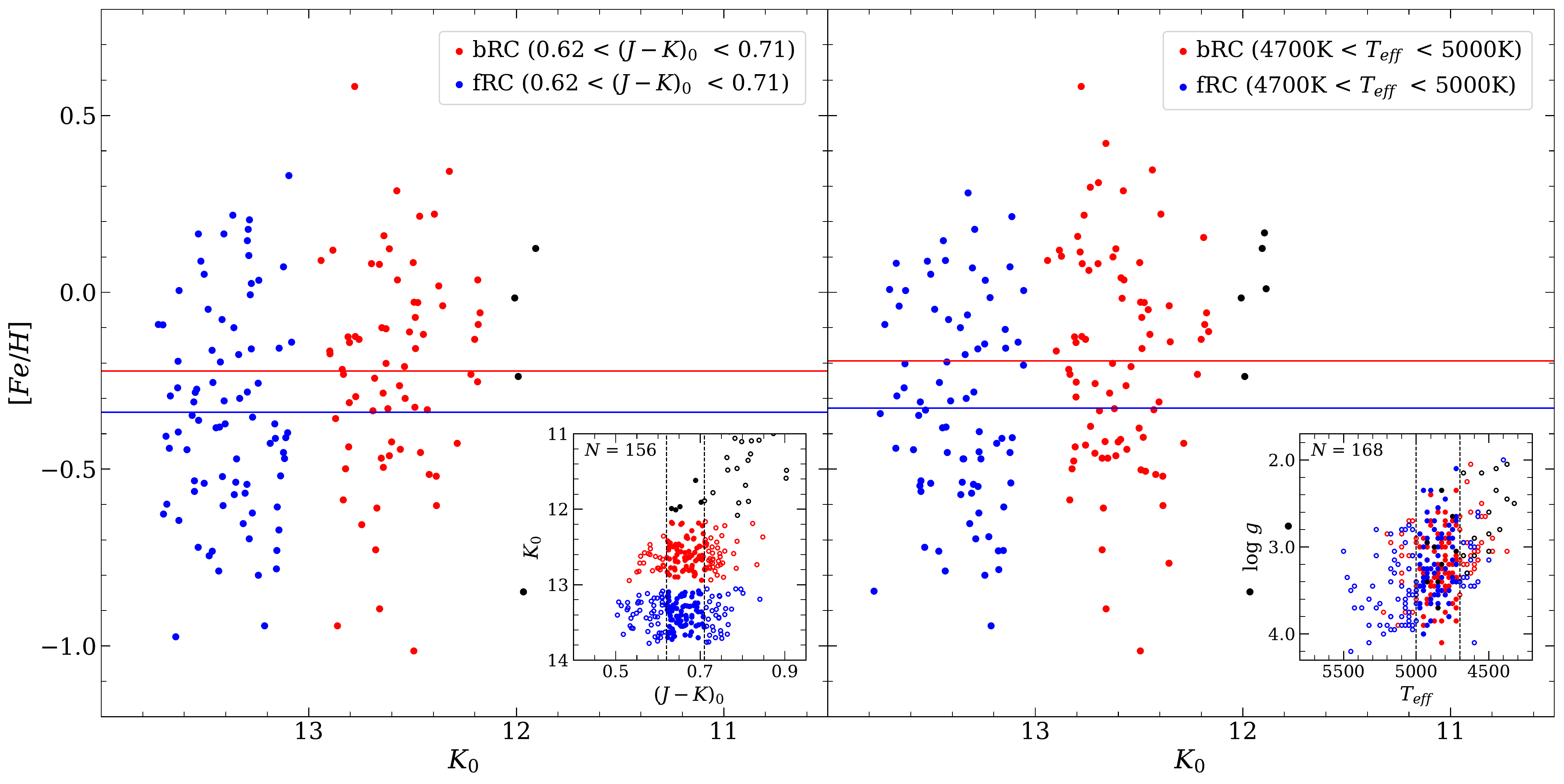}
\figcaption{
The comparison of [Fe/H] between the stars in the bRC and fRC groups within a limited range of $(J-K)_{0}$ ($left$) and $T$\textsubscript{eff} ($right$). The blue and red horizontal lines indicate the mean [Fe/H] values for selected subsamples in each group. The insets show selection criteria on the CMD and log~$g$ - $T$\textsubscript{eff} diagram, respectively, where the filled and open symbols denote selected and rejected sample stars. In both cases, the stars in the bRC group are more enhanced in [Fe/H] than those in the fRC group, which is identical to the result obtained from all samples (see Figure~\ref{fig:FeH}). 
\label{fig:FeH_teff}
}
\end{figure*}

The difference in metallicity between the stars in the bRC and fRC groups has already been reported in a field at ($l$, $b$) = (0$\degree$, -10$\degree$) by \citet{Uttenthaler2012}. The mean metallicity was estimated to be [M/H] = $-$0.18 for the bRC, and $-$0.41 for the fRC, respectively, with a difference of 0.23 dex, which is in agreement with our estimate. They attributed this difference to be rather due to a selection bias, as redder stars (metal-rich) are more selected from the bRC regime while bluer stars (metal-poor) are preferentially selected from the fRC regime. In order to examine this hypothesis, we compare the metallicity of stars selected from the narrow ranges of color and temperature so that they have almost the same color and temperature. Figure~\ref{fig:FeH_teff} shows the metallicity difference between the stars in bRC and fRC groups, within a limited range of color (0.62 $<$ $(J-K)_{0}$ $<$ 0.71; left panel) and temperature (4700K $<$ $T$\textsubscript{eff} $<$ 5000; right panel), respectively. These criteria are devised to include $\sim$ 50\% of total sample stars from the mean value of color and temperature. In both cases, stars in the bRC group are more enhanced in the mean metallicity than those in the fRC group. The difference in [Fe/H] is estimated to be 0.117 $\pm$ 0.048 dex from the limited color, and 0.134 $\pm$ 0.047 dex from the limited temperature subsamples, with the $p$-values of 0.01 and 0.03, respectively, from the KS-test. The metallicity variation is still meaningful at the $p$-values less than 0.07 level, unless we apply very narrow color (0.64 $<$ $(J-K)_{0}$ $<$ 0.69) and temperature (4750K $<$ $T$\textsubscript{eff} $<$ 4950K) criteria. This reassures that the metallicity difference between the stars in bRC and fRC groups is not due to the selection bias in color or temperature, but due to the chemical composition. 

\begin{figure}
\centering
\includegraphics[width=0.48\textwidth]{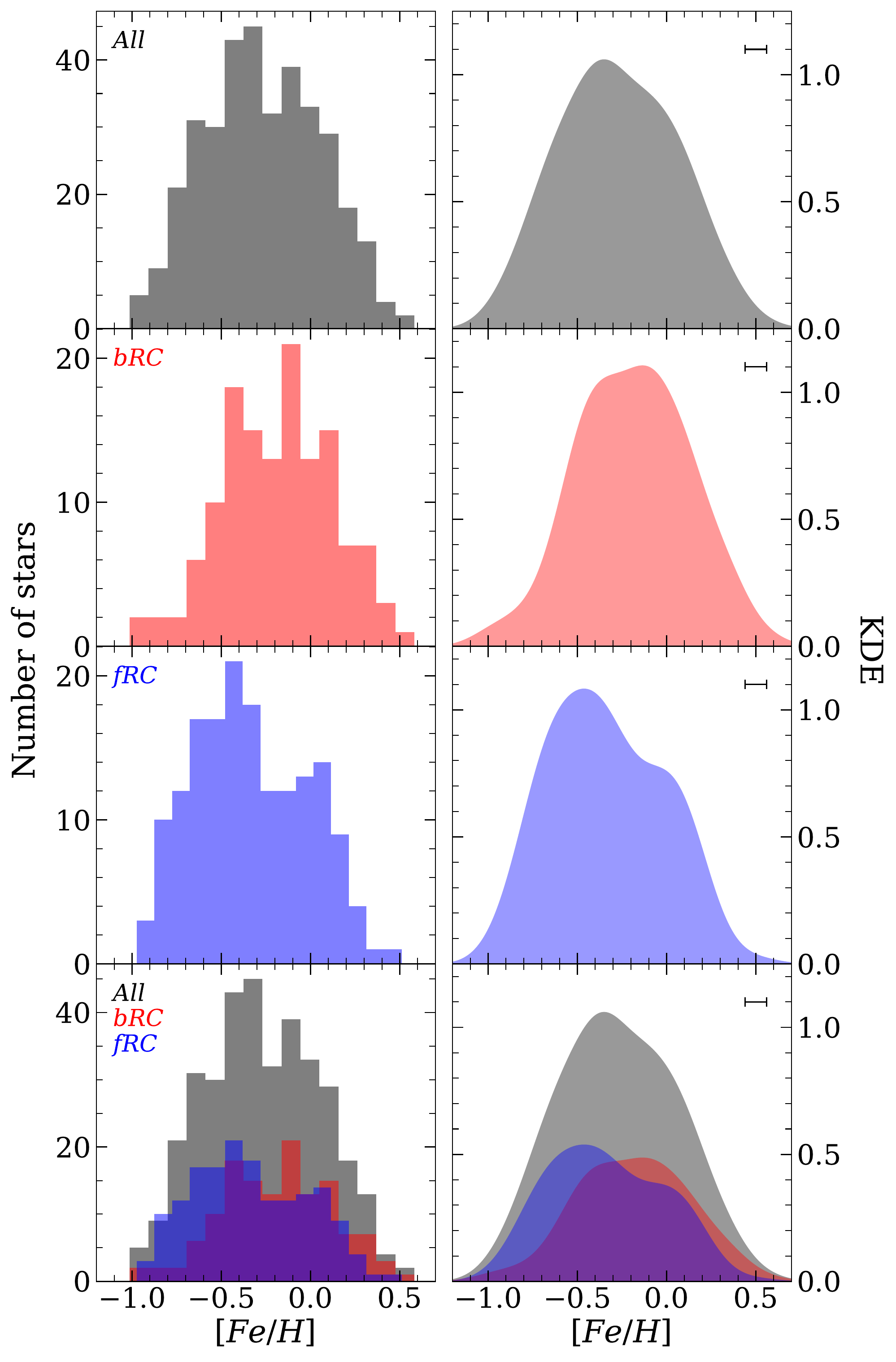}
\figcaption{
Metallicity distribution and KDE for all samples (grey), and stars in each group, bRC (red) and fRC (blue). The bRC and fRC groups show contrasting distributions in [Fe/H], where the bRC is generally more metal-rich. In addition, the fRC group has two distinct peaks, whereas it is not pronounced in the bRC group. Note that the bandwidth for KDE is estimated by Silverman's rule of thumb (the horizontal bar in the upper right corner of the right panels). 
\label{fig:FeH_hist}
}
\end{figure}

In Figure~\ref{fig:FeH_hist}, we plot the metallicity distribution functions and kernel density estimators (KDE) for stars in each group, respectively. As is expected from the difference in mean metallicity between the bRC and fRC groups, the two groups show contrasting distributions in [Fe/H], in the sense that the stars in the bRC are generally more metal-rich than those in the fRC. A closer inspection of KDE diagram shows a possible bimodal distribution of [Fe/H] from the combined sample (the top right panel of Figure~\ref{fig:FeH_hist}). Therefore, we performed Gaussian Mixture Modeling (GMM; \citealt{Muratov2010}) test in order to investigate whether the distribution of stars is rather uni- or bimodal. According to \citet{Ashman1994} and \citet{Muratov2010}, $D$ $>$ 2 and $kurt$ $<$ 0 are required to confirm a clean separation between two Gaussians, where the values $D$ and $kurt$ are the separation of the means relative to their widths and the kurtosis of the distribution, respectively. Through the GMM test, we confirm the presence of two populations ($D$ = 2.20 $\pm$ 0.24 with $kurt$ = -0.668) with mean values of -0.48 dex and 0.00 dex in [Fe/H]. The difference between the two populations is close to that expected from genuine RC stars in our analysis ($\sim$ 0.55 dex; see above). In addition, stars in either group show a bimodal distribution of metallicity ($D$ = 2.78 $\pm$ 0.39 with $kurt$ = -0.832 for the fRC; $D$ = 1.72 $\pm$ 0.89 with $kurt$ = -0.341 for the bRC), although it is less pronounced in the bRC. These results are consistent with the earlier findings of two or more stellar components with different metallicity in the MW bulge \citep[e.g.,][]{Ness2013a, Gonzalez2015, Johnson2020}. It is important to note that the MW bulge might host more than three components with different metallicity, while we only use two Gaussian models for the analysis. It is also possible that the bulge RC stars do not show Gaussian distributions of metallicity. In these cases, a more detailed interpretation for each stellar component in the bulge will be required \citep[e.g.,][]{Kunder2012,Ness2013a,McWilliam2016}.

\subsection{metal-poor vs. metal-rich components} 

\begin{figure}
\centering
\includegraphics[width=0.47\textwidth]{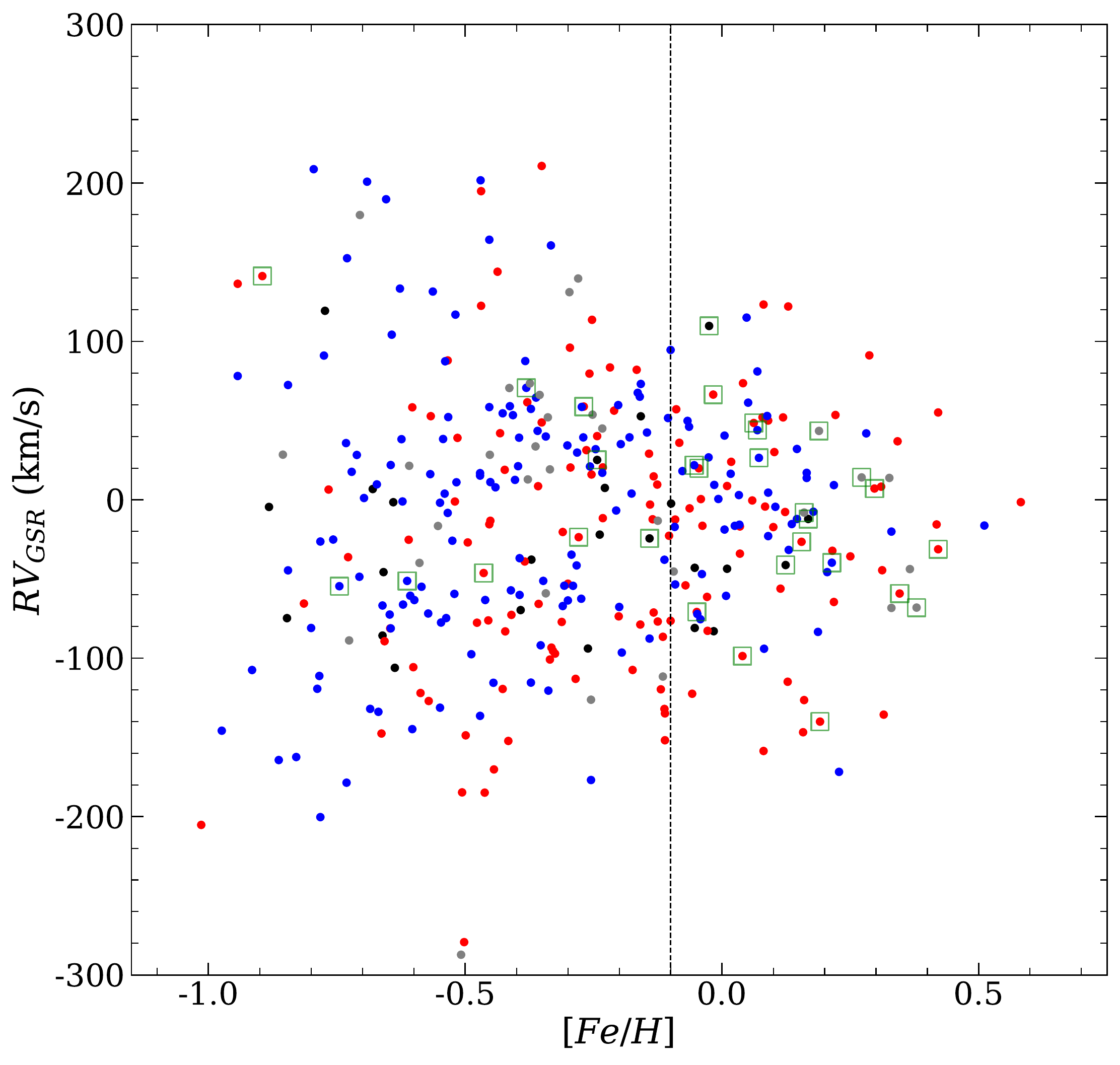}
\figcaption{
RVs of stars with respect to the Galactic Standard of Rest (GSR). The symbols are the same as in Figure~\ref{fig:CMD}. The vertical dashed line divides the metal-poor and metal-rich components at [Fe/H] = $-$0.1 dex. The metal-poor component shows a higher velocity dispersion than the metal-rich component ($\sigma$ $\sim$ 90 km/s for metal-poor and $\sigma$ $\sim$ 60 km/s for metal-rich stars). In addition, green squares indicate Na-rich stars (see Section~\ref{sec:other} and Figure~\ref{fig:trend}), and these stars do not show any specific trend of RVs.
\label{fig:RV}
}
\end{figure}

\begin{figure}
\centering
\includegraphics[width=0.47\textwidth]{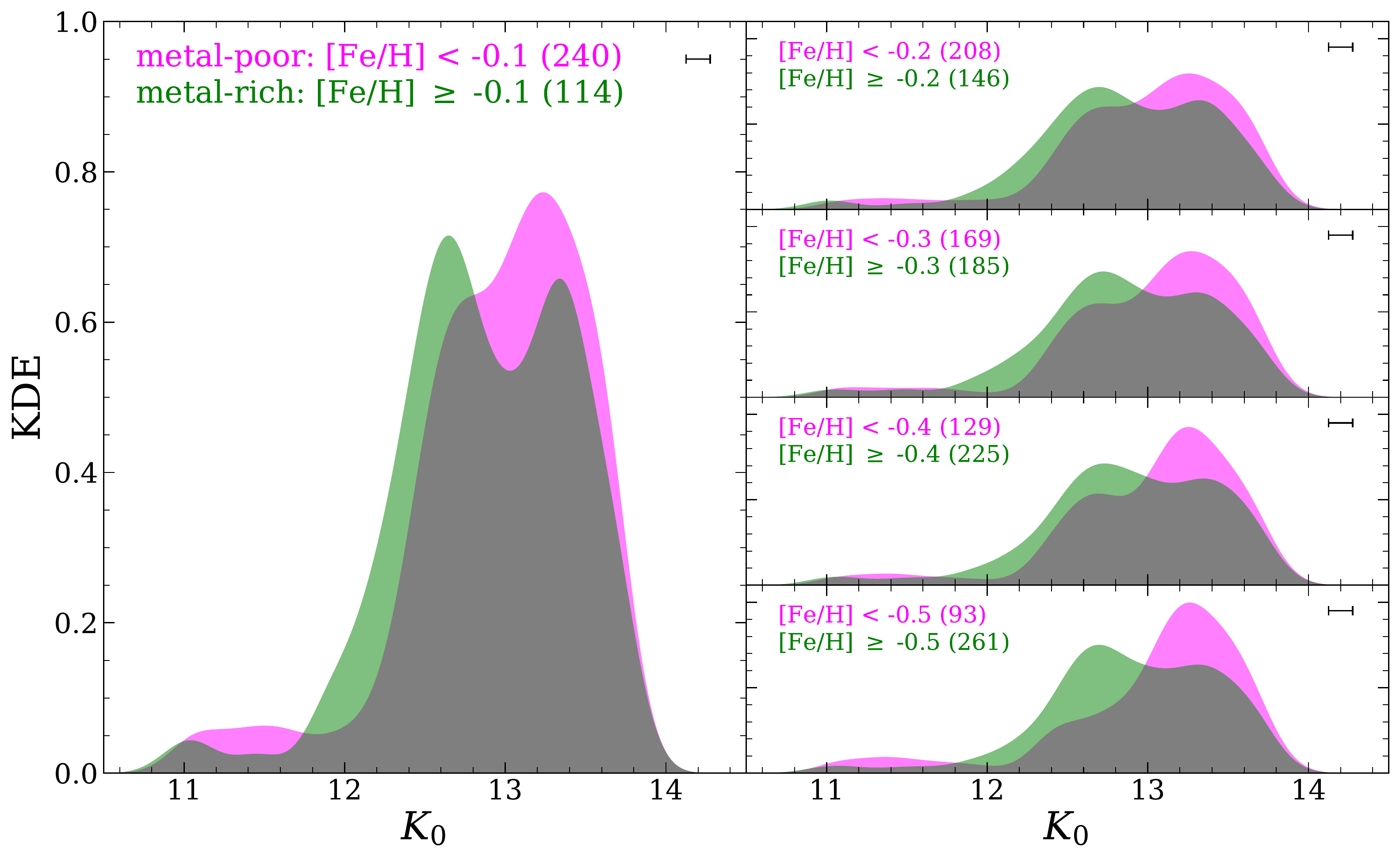}
\figcaption{
Luminosity functions for the metal-poor (purple) and metal-rich (green) components according to different separation criteria at [Fe/H] = $-$0.1, $-$0.2, $-$0.3, $-$0.4 and $-$0.5 dex. The criterion and number of stars in each component are written in each panel. The double RC feature is observed not only in the metal-rich component but also in the metal-poor one in every panel. 
\label{fig:mp_mr}
}
\end{figure}

Recent studies reported that the MW bulge hosts metal-poor and metal-rich stellar components with different kinematic characteristics \citep[e.g.,][]{Ness2013a, Zoccali2017}. The metal-rich component has a steeper gradient of velocity dispersion with Galactic latitude compared to the metal-poor component, and therefore the metal-rich component shows a lower velocity dispersion than the metal-poor component in the high-latitude field at $b$ = -8.5$\degree$. These kinematic properties have long been studied as an evidence of distinct populations in the bulge \citep{Rich1990,Minniti1992,Zoccali2008,Babusiaux2010,Johnson2011,Ness2013b}.
Our observation also demonstrates two stellar components in the distribution of RV. In Figure~\ref{fig:RV}, metal-poor stars show a higher velocity dispersion than metal-rich stars ($\sigma$ $\sim$ 90 km/s for stars with [Fe/H] $<$ -0.1 and $\sigma$ $\sim$ 60 km/s for stars with [Fe/H] $\geq$ -0.1), which is quantitatively identical to that illustrated in Figures~10 and 11 of \citet{Zoccali2017} for their {\em p0m8} field. It is important to note that there is no difference in velocity dispersion or mean RV between the stars in the bRC and fRC groups ($\sigma$ $\sim$ 85 km/s for both bRC and fRC), which is also compatible with the result of \citet{Zoccali2017}. 

Furthermore, the inner MW spheroid has often been reported to be metal-poor \citep{Tumlinson2010,Ness2013a,Koch2016,Kunder2016,Savino2020}, while the metal-rich stars rather tend to follow the boxy shape which is commonly associated with the X-shaped structure \citep{Rojas-Arriagada2014, Zoccali2017}. In this regard, if the double RC is originated from the different distance of stars in the X-shaped structure, it would be observed mainly in the metal-rich component. In order to examine this, we divide all samples, including stars in the twilight zone, into metal-poor and metal-rich components at [Fe/H] $\sim$ -0.1 (vertical dashed line in Figure~\ref{fig:RV}), and then draw the luminosity function for each component in the left panel of Figure~\ref{fig:mp_mr}. Both metal-rich and metal-poor components show a distinct double RC feature with opposite trends on the bright and faint RC regimes. The metal-poor stars are more dominant in the fRC than the bRC, while this trend is reversed in the metal-rich component. Note that a similar trend has already been reported by \citet[][see their Figure~20]{Ness2013a} without a detailed description of the metal-poor component. Since the fraction of genuine RC star is small ($\sim$ 27\%) in the RC magnitude range, this trend is consistent with what we would expect from the multiple population scenario as metal-poor RC stars (G1) are embedded in metal-poor and metal-rich RGB stars (G1+G2\textsuperscript{+}) at the fRC regime, whereas the bRC regime contains metal-rich RC stars (G2\textsuperscript{+}) together with G1+G2\textsuperscript{+} RGB stars. In addition, we confirm that the double RC feature is prominent in both the metal-rich and metal-poor components, as witnessed through the GMM test ($D$ = 3.26 $\pm$ 0.36 for the metal-rich, and 2.72 $\pm$ 0.33 for the metal-poor component). It appears, therefore, that the presence of a double RC in both the metal-rich and metal-poor components, as well as the different fractions of the bright and faint RCs between the two components, corresponds with the multiple population scenario. In addition, the dominance of fRC in the metal-poor spheroidal  component is inexplicable without the multiple population phenomenon because stars closer to us should be more prominently observed in the bRC regime. As shown in right panels of Figure~\ref{fig:mp_mr}, the double RC feature still appears in both the metal-rich and metal-poor components with high probability (GMM test $D$ $>$ 2.9 and $kurt$ $\sim$ -1.0) when we apply a lower criterion at [Fe/H] = $-$0.2, $-$0.3, $-$0.4 and $-$0.5 dex, respectively. One possible caveat of this approach is a shortage of stars in the `twilight zone' by selection bias, which can mimic the bimodal luminosity distribution. However, we emphasize that the target stars were randomly selected in the magnitude range of $K$ from 12.0 to 14.0.


\section{Other chemical properties of RC stars} \label{sec:other}
\begin{figure}
\centering
\includegraphics[width=0.45\textwidth]{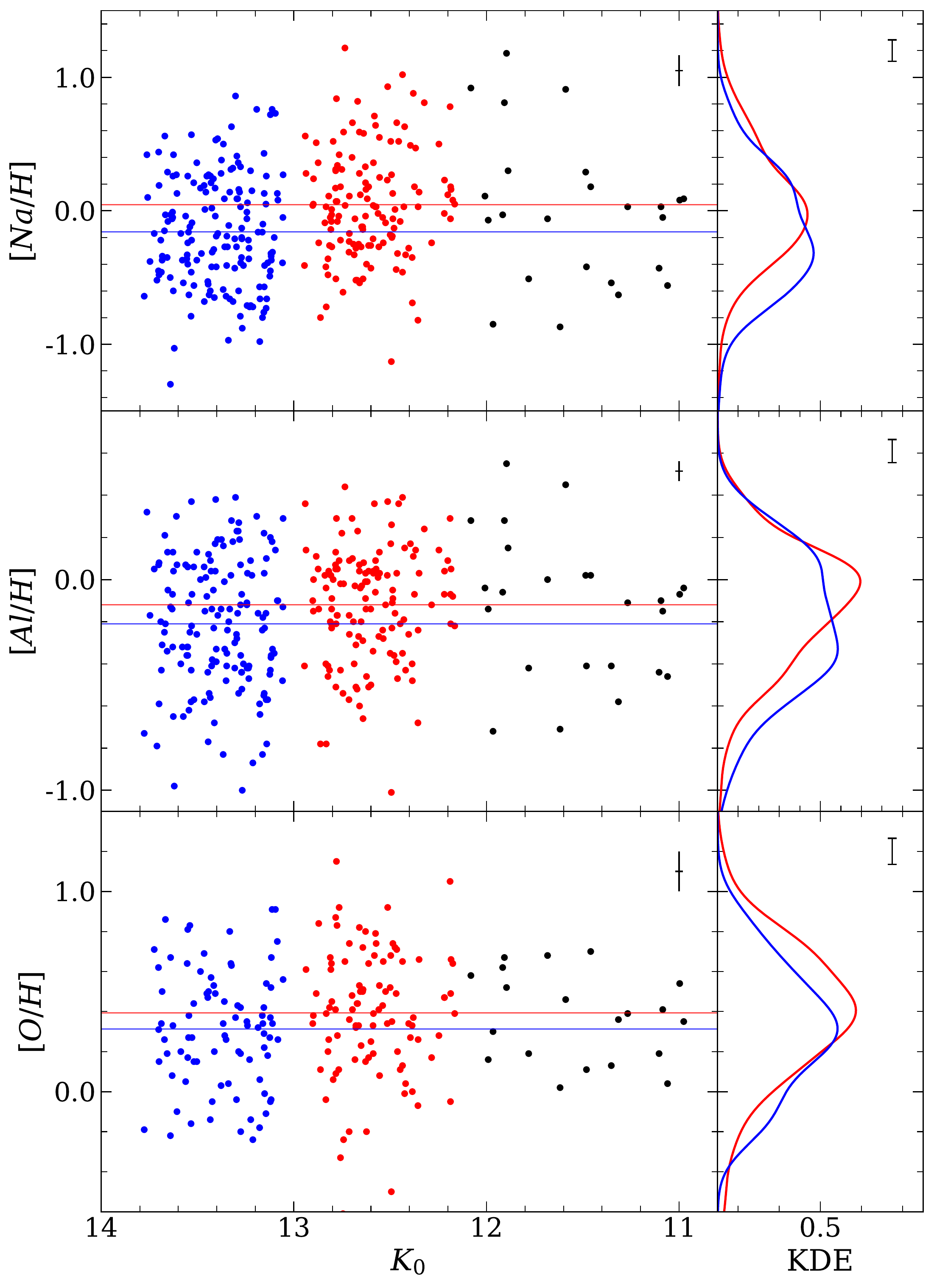}
\figcaption{
Measured [Na/H], [Al/H], and [O/H] abundances as a function of $K_{0}$ magnitude (left panels) and the KDE histogram (right panels) for stars in the bRC and fRC groups. The symbols are the same as in Figure~\ref{fig:FeH}. The stars in the bRC and fRC groups show differences in [Na, Al, O/H] abundances not only in the mean value but also in the shape of the distribution, where the bRC is more enhanced. The typical measurement error for each abundance is shown in the upper-right corner of the left panels and the vertical bars in the right panels indicate the bandwidth for KDE.
\label{fig:NaH}
}
\end{figure}

\begin{figure}
\centering
\includegraphics[width=0.45\textwidth]{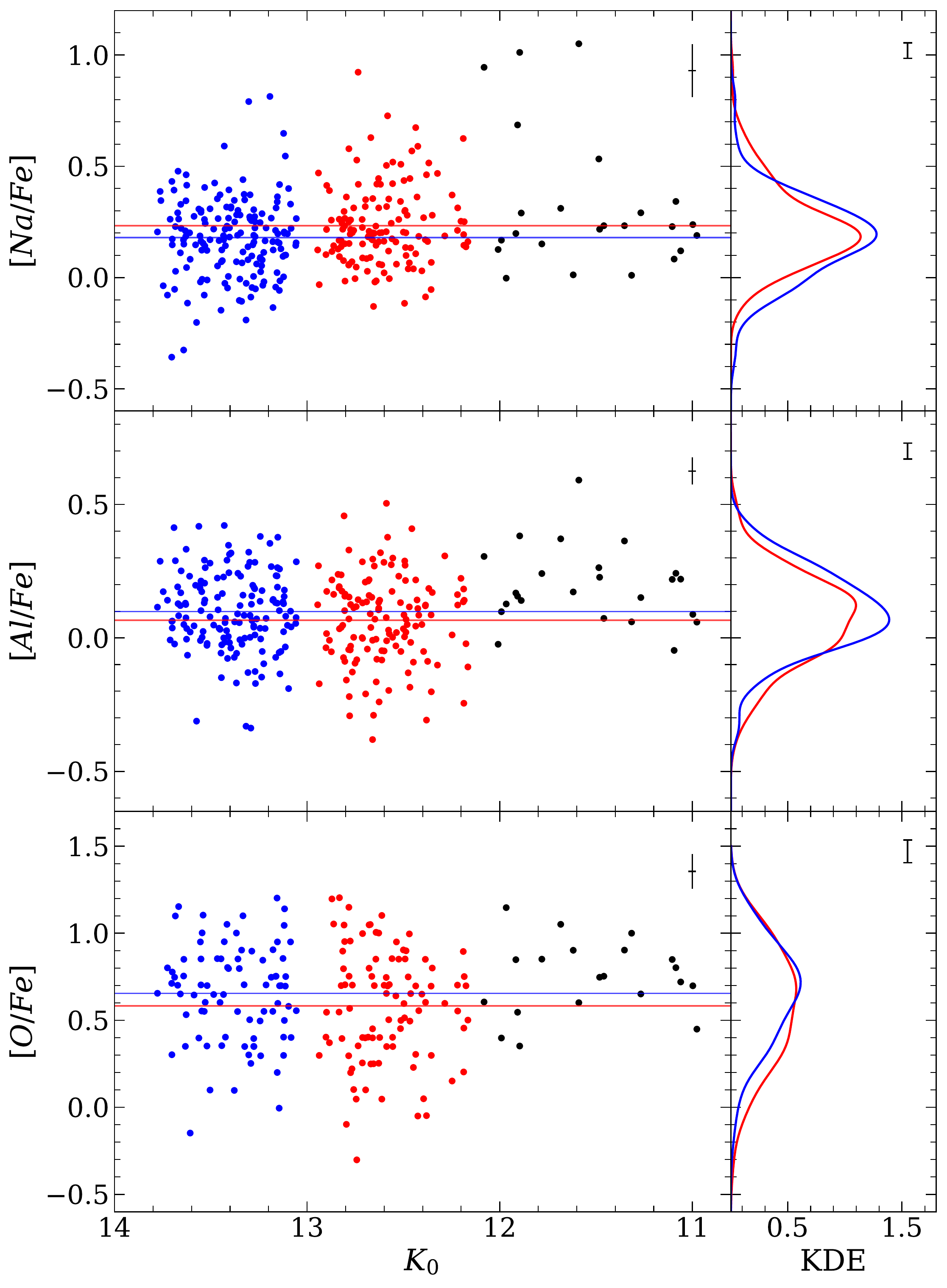}
\figcaption{
Same as Figure~\ref{fig:NaH}, but for [Na/Fe], [Al/Fe], and [O/Fe] abundance ratios. The stars in the bRC group are slightly more enhanced in [Na/Fe] but more depleted in [Al/Fe] and [O/Fe] than those in the fRC group. These chemical patterns are similarly observed in Terzan~5 between metal-poor and metal-rich populations.
\label{fig:NaFe}
}
\end{figure}

The abundances of Na, O, Al, and Mg elements are crucial tracers of multiple stellar populations since the Na-O and Al-Mg anticorrelations are the common characteristics of MW GCs with multiple populations \citep[see, e.g.,][]{Carretta2009a, Carretta2009b, Bastian2018}. In order to compare the abundances of these elements between the stars in the bRC and fRC groups, we plotted [Na, Al, O/H] and [Na, Al, O/Fe] abundances as a function of magnitude in Figures~\ref{fig:NaH} and \ref{fig:NaFe}, together with generalized histograms for each group. As described above, the abundance of Mg could not be derived due to the strong Ca~\textsubscript{I} autoionization in the region of Mg~\textsubscript{I} absorption lines. 

\begin{figure}
\centering
\includegraphics[width=0.45\textwidth]{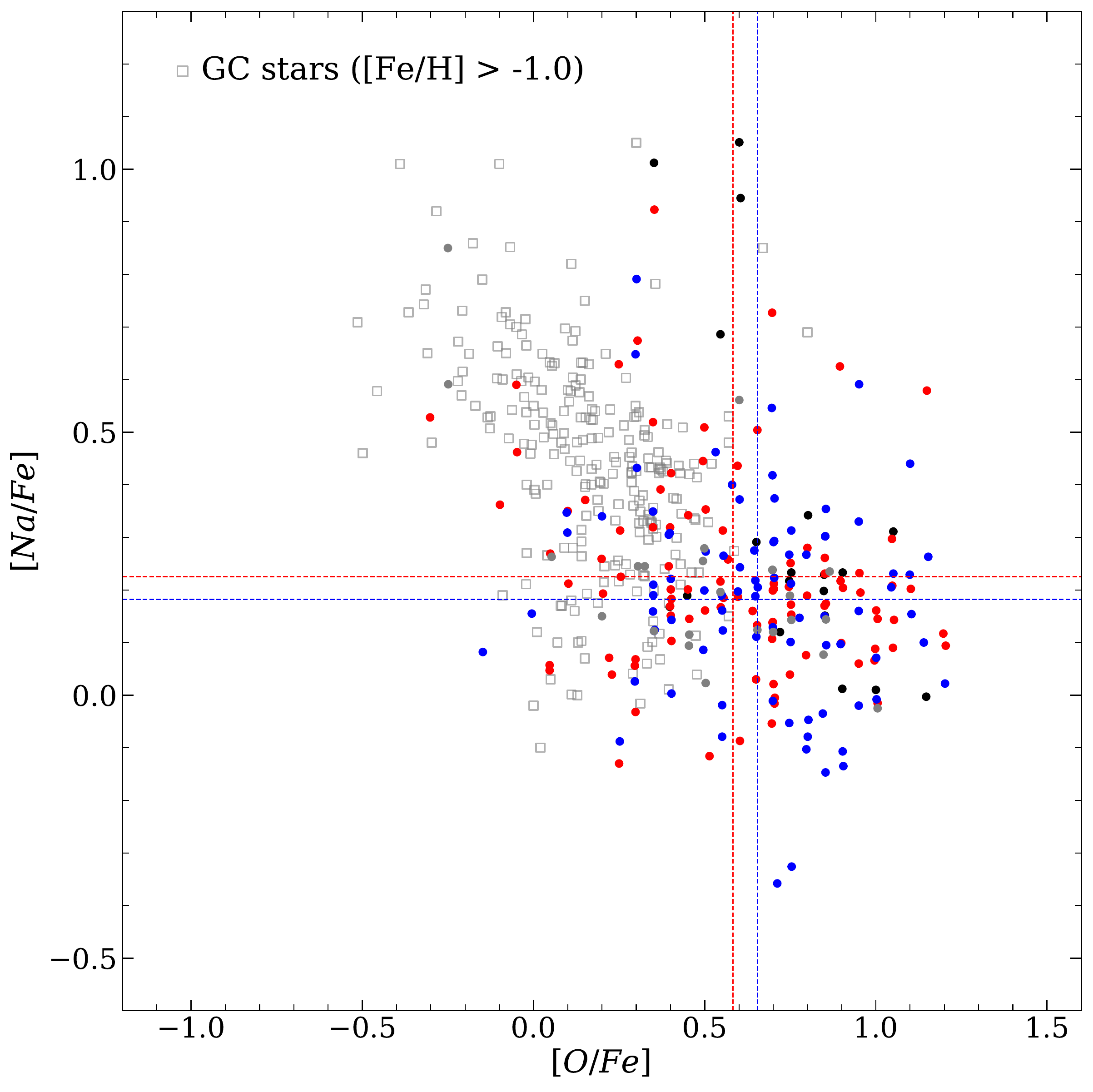}
\figcaption{Observed bulge stars on the Na-O plane, together with star in metal-rich GCs (open squares; [Fe/H] $>$ -1.0 dex, data from \citealt{Gratton2006,Gratton2007,Carretta2009a,Carretta2009b,Munoz2017,Munoz2018}). The vertical and horizontal dashed lines represent mean values of [O/Fe] and [Na/Fe] for the bRC (red) and the fRC (blue) groups, respectively. Although the stars in the bRC group are slightly more enhanced in [Na/Fe] and depleted in [O/Fe] than stars in the fRC group on average, the distribution pattern of the RC stars differs from typical GC stars.
\label{fig:NaO}
}
\end{figure}

As shown in Figure~\ref{fig:NaH}, stars in the bRC group are more enhanced in [Na/H], [Al/H], and [O/H] abundances than those in the fRC group. The mean differences between the two groups are 0.202 $\pm$ 0.048 dex in [Na/H], 0.091 $\pm$ 0.034 dex in [Al/H], and 0.080 $\pm$ 0.045 dex in [O/H], respectively. The $p$-values of the KS-test, which indicates the probability that two groups are drawn from the same distribution, are small for [Na/H] ($p$ = 0.0002) and [Al/H] ($p$ = 0.0097), but somewhat larger for [O/H] ($p$ = 0.1243). The enhancement of these abundances in stars belonging to the bRC group is consistent with the behavior expected from the enhancement in [Fe/H] (see Section~\ref{sec:metal}). The distribution patterns of [Na/H] and [Al/H] abundances for the bRC and fRC groups are particularly similar to that of [Fe/H] (see Figure~\ref{fig:FeH_hist} and the right panels of Figure~\ref{fig:NaH}). This result, therefore, indicates that stars in the bRC group are more enhanced in the overall metallicity than those in the fRC group, suggesting a difference in chemical compositions between the genuine bright and faint RC stars. 

In Figure~\ref{fig:NaFe}, we also find small differences in the [Na, Al, O/Fe] abundance ratios between the stars in the bright and faint RC groups, although it is not as clear as in the case of [Na, Al, O/H] abundances. In particular, unlike Figure~\ref{fig:NaH}, stars in the bRC group are more enhanced in [Na/Fe] but appear to be depleted in [Al/Fe] and [O/Fe] than those in the fRC group. The mean differences are 0.053 $\pm$ 0.021 dex, 0.032 $\pm$ 0.018 dex, and 0.071 $\pm$ 0.045 dex in [Na/Fe], [Al/Fe], and [O/Fe], respectively, which are marginally significant at $p$-values of 0.22, 0.18 and 0.23. When the relative fraction of RC stars is taken into account (27\%; see Section~\ref{sec:metal}), the difference in [Na/Fe] between the genuine RC stars would correspond to $\Delta$[Na/Fe] $\sim$ 0.20 dex, which is comparable to that expected from our chemical evolution model for the bulge stars ($\Delta$[Na/Fe] = 0.2 $\sim$ 0.3 dex; \citealt{Kim2018, Lee2019})\footnote{The previous study by  \citet{Lee2019} noted a clear separation of the two groups according to Na abundance amongst bright RGB stars in the outer bulge. The apparent lack of such a distinct difference between the two groups in this study may be due to a larger uncertainty on abundances of relatively faint sample stars.}. The overall chemical patterns, however, are not identical to those observed in typical GCs, where the later-generation stars are enhanced in [Na, Al/Fe] and depleted in [O, Mg/Fe] than the first-generation stars at given metallicity, although the trend of [Na, Al O/Fe] between the two RCs is less clear. Figure~\ref{fig:NaO} shows the comparison of stars in this study with stars in metal-rich GCs ([Fe/H] > -1.0) on the Na-O diagram. The stars used in this study have a different distribution from stars in GCs. Although the bRC group is slightly more enhanced in [Na/Fe] and depleted in [O/Fe] than the fRC group, the [Na/Fe] variation of RC stars is smaller than that of GC stars. This discrepancy might imply the different chemical evolution between stars in the bulge and typical GCs. We note, however, that even though we employ only metal-rich GC stars for the comparison, the majority of stars are still far metal-poor ([Fe/H] $<$ -0.5) than stars in the bulge. Because the relatively small [Na/Fe] variation is expected from the chemical evolution model for metal-rich bulge stars and the O-depletion is indistinct in some metal-rich GCs, such as NGC~6121 and 47 Tuc \citep[see][]{Kim2018, Lee2019}, the direct comparison of bulge stars with similarly metal-rich GCs on the Na-O plane would require further spectroscopic observations for such GCs in the bulge.
\begin{figure*}
\centering
\includegraphics[width=0.83\textwidth]{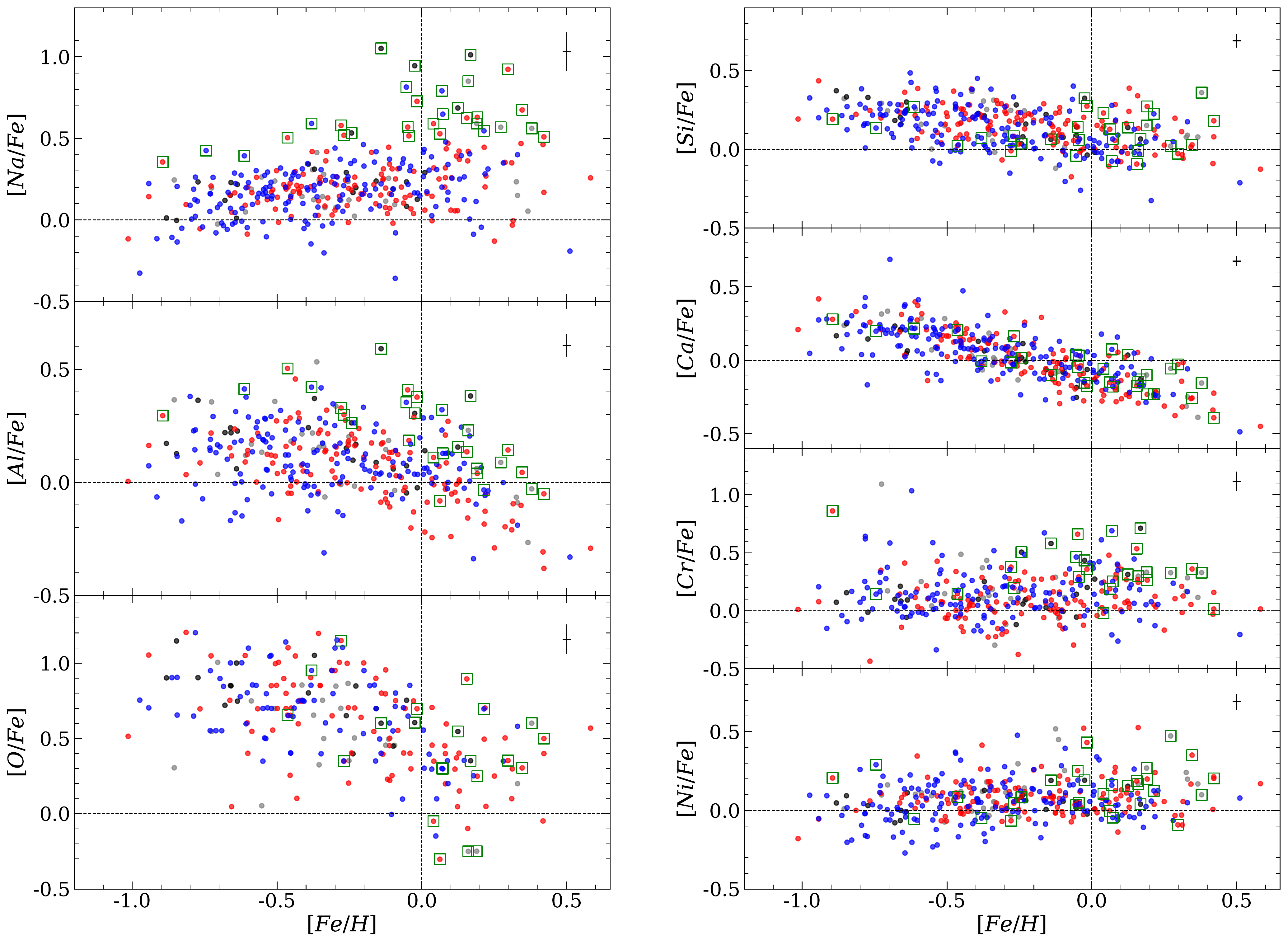}
\figcaption{
[X/Fe] abundances distribution as a function of [Fe/H] for the elements measured in this study. The symbols are the same as in Figure~\ref{fig:CMD}, and horizontal and vertical dashed lines indicate the solar level. [Na/Fe] abundances increase with increasing [Fe/H], while [Al/Fe], [O/Fe], [Si/Fe], and [Ca/Fe] abundances decrease. Green squares indicate Na-rich stars, which are also enhanced in [Al/Fe], but not in [O/Fe]. A total of 30 Na-rich stars (7 in fRC and 14 in bRC) would be G3 stars in the bulge (see text). The typical measurement error is shown in the upper-right corner of each panel. 
\label{fig:trend}
}
\end{figure*}

Interestingly, these chemical patterns between the bugle bRC and fRC populations are similarly observed in Terzan~5 between its metal-rich and metal-poor populations. \citet{Origlia2011} reported that the metal-rich population of Terzan~5 is more depleted in both [Al/Fe] and [O/Fe] than the metal-poor population, unlike other typical GCs. They explained this as due to the chemical enrichment by both type Ia and II supernovae (SNe) on a longer timescale to the formation of the
later, metal-rich population. In contrast, the abundance variations in typical GCs are generally considered to be due to pollution from massive asymptotic giant branch (AGB)  or fast rotating massive stars, without the need to invoke a large age difference between subpopulations \citep[see, e.g.,][]{Bastian2018}. In our observations, the stars belonging in the bRC group are relatively metal-rich and more depleted in [Al/Fe] and [O/Fe] abundances than stars in the fRC group. In the case of the $\alpha$-elements (Ca and Si), the mean value of the metal-poor fRC stars is larger than that of the metal-rich bRC stars, which is also in agreement with that observed in Terzan~5. In addition, the difference in [Na/Fe] ($\sim$ 0.20 dex), expected within genuine RC stars, is comparable to that of Terzan~5 ($\Delta$[Na/Fe] $\sim$ 0.25 dex; \citealt{Schiavon2017a}), although this is based on a small sample. These similarities suggest that the bRC (fRC) stars in the bulge would correspond to the metal-rich (metal-poor) population of Terzan~5, although some unique properties of Terzan~5, such as three abundance groups and large age distribution, should be further addressed. In this regard, the chemical properties of the two RCs, together with the remarkably analogous $\Delta$[Fe/H] to Terzan~5, again suggest that the Terzan~5-like stellar systems could be the building blocks to form the double RC feature in the bulge. 

\begin{figure}
\centering
\includegraphics[width=0.48\textwidth]{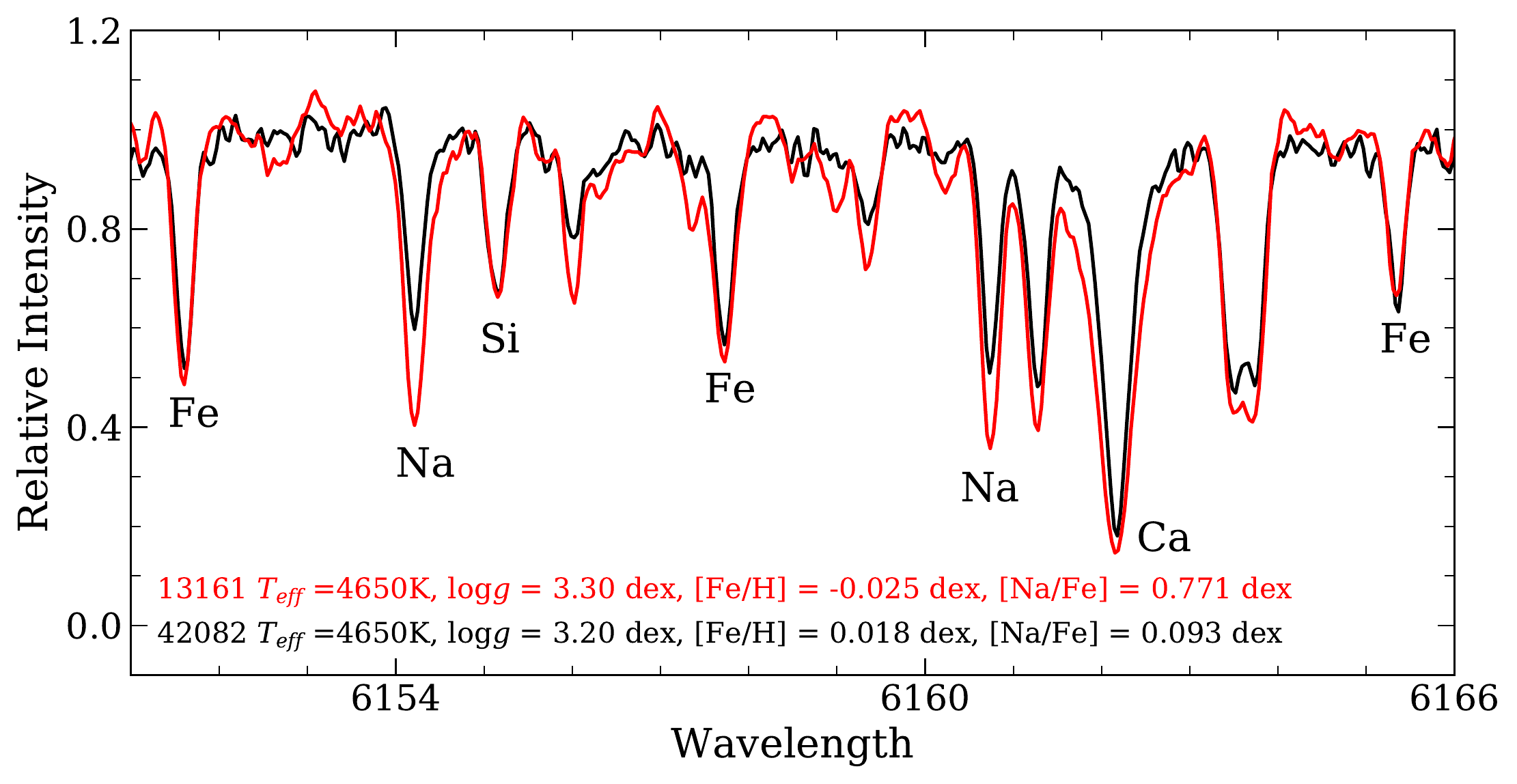}
\figcaption{
Example of the spectra for the Na-rich (red) and Na-normal (black) stars. Two stars have a similar $T$\textsubscript{eff}, log~$g$, and [Fe/H], but they show a clear difference in [Na/Fe] abundance. 
\label{fig:Na_spectra}
}
\end{figure}

\begin{figure}
\centering
\includegraphics[width=0.48\textwidth]{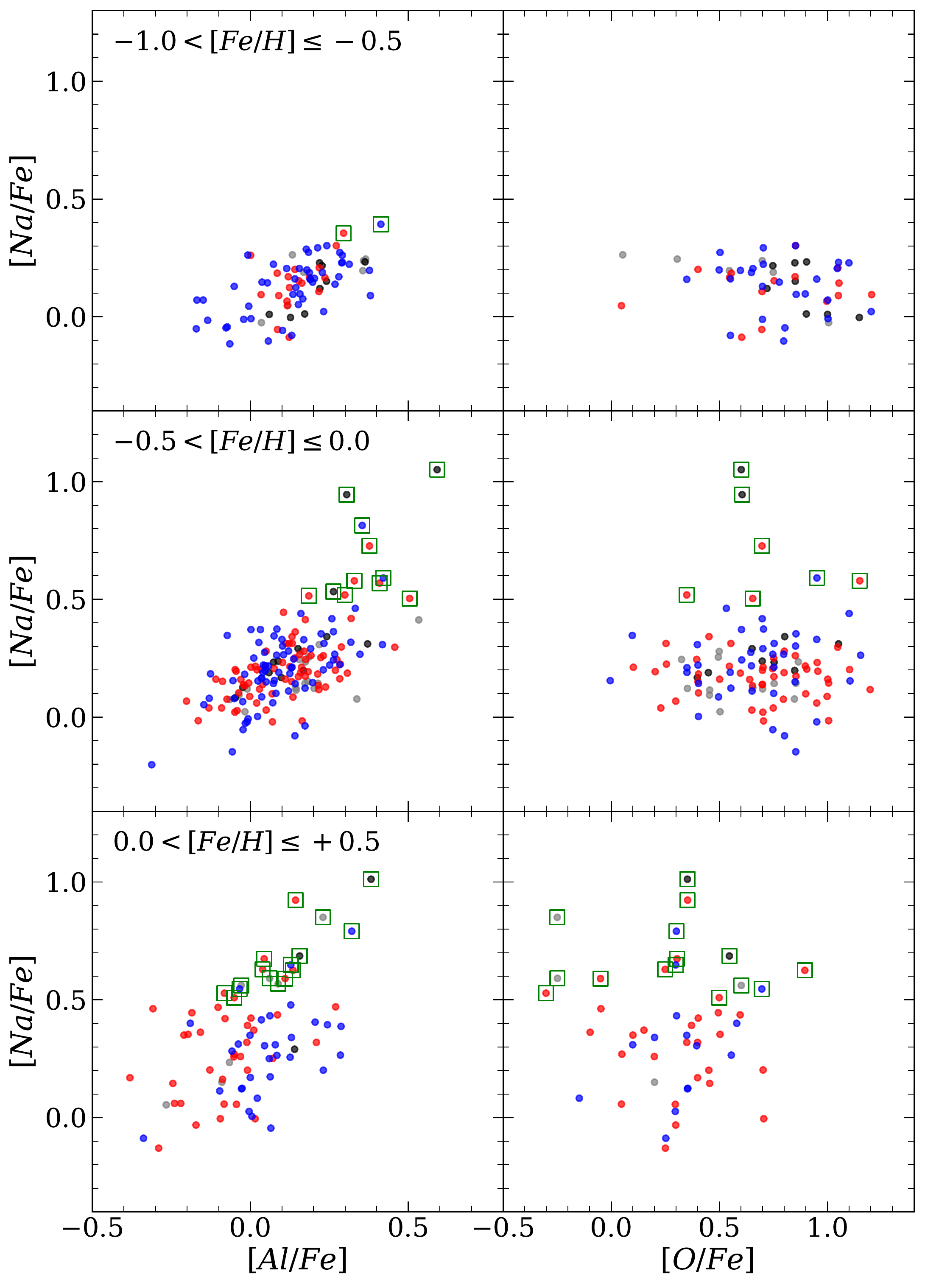}
\figcaption{
[Na/Fe] abundance as a function of [Al/Fe] and [O/Fe] abundances for stars in three metallicity groups, respectively. The symbols are the same as in Figure~\ref{fig:trend}. The Na-rich stars are also enhanced in [Al/Fe], while no trends are observed in [O/Fe].
\label{fig:Na_Al}
}
\end{figure}

Figure~\ref{fig:trend} shows the chemical abundances of our sample stars as a function of [Fe/H] for all elements measured in this study. These chemical abundances follow the general trends of the MW bulge field reported by other spectroscopy surveys \citep[e.g.,][]{Johnson2014, Koch2016, Zasowski2019}. Note, however, that our measurement of Ca abundance is lower than that of other studies, while Na and O abundances are marginally enhanced. Interestingly, we find 30 stars which have high [Na/Fe] abundance ratios\footnote{It appears that the high [Na/Fe] abundances are not due to the measurement errors. The mean measurement error for these Na-rich stars is 0.20 dex, which is comparable to that for all samples ($\sim$ 0.24 dex), and only 7 out of 30 stars have relatively large measurement error (> 0.30 dex). In addition, we could not find any erroneous feature from visual inspection of Na lines for each spectrum (see, e.g., Figure~\ref{fig:Na_spectra}).} (green squares in Figure~\ref{fig:trend}; [Na/Fe] > 0.35 for stars with [Fe/H] < -0.5; [Na/Fe] > 0.5 for stars with [Fe/H] $\ge$ -0.5). These Na-rich stars have already been shown in earlier studies for the bulge \citep[e.g.,][]{Lecureur2007, Johnson2014}. Figure~\ref{fig:Na_spectra} shows an example spectrum for the Na-rich stars, together with that for Na-normal star having similar $T$\textsubscript{eff}, log~$g$, and [Fe/H] with the Na-rich star. Two spectra show a clear difference in the strength of two Na lines at 6154~{\AA} and 6160~{\AA}. These stronger Na lines are similarly detected in other spectra of Na-rich stars. We also plot the [Na/Fe] abundances as a function of [Al/Fe] and [O/Fe] for three metallicity groups in Figure~\ref{fig:Na_Al}. In each group, the Na-rich stars are also enhanced in [Al/Fe], while no specific tendency is observed in [O/Fe]. The Al-enhancement of Na-rich stars compared to the other stars of similar metallicity is already shown in Figure~\ref{fig:trend}. In addition, these stars seem to be slightly more enhanced in [Cr/Fe] than the majority of stars, but they are embedded in the mainstream of stars on the [Si, Ca, Ni/Fe] abundances. We note that there is no kinematic difference of the Na-rich stars seen in the RV distribution (see Figure~\ref{fig:RV}).

These unique chemical abundance patterns are consistent with those reported from the Na-rich stars (G2\textsuperscript{+}) on the bright RGB by \citet{Lee2019}. Recent studies demonstrated that most GCs host more than three subpopulations, where the third or later generation are more enhanced in Na and Al abundances \citep[see, e.g.,][]{Carretta2015,Johnson2019}. In this regard, these Na-rich stars might be considered as G3 (third generation) stars in the bulge. Since G3 is a chemically enhanced stellar population, we expect that those stars would be placed on the bRC regime together with G2 stars. Thus, the genuine bRC stars would be composed of G2 and G3 stars, while G1 stars on the fRC, together with background RGB stars (G1+G2+G3) in both regimes. As shown in Figure~\ref{fig:trend}, among a total of 30 Na-rich stars, 7 are in the fRC group, and 14 are in the bRC group. When we assume that 7 Na-rich stars in the fRC are G3 RGB stars, the fraction of G3 stars in the fRC regime is about 5-6\% of total RGB stars. This fraction indicates that about 10--11 Na-rich stars are required in the bRC group including G3 RC and RGB stars. Despite the small sample size, the number of Na-rich stars in the bRC group is roughly in agreement with this estimate. We also note that the Na-rich stars are more metal-rich (mean [Fe/H] = $-$0.04 $\pm$ 0.06 dex) compared to other stars as expected from chemically enhanced G3 population. Therefore, our finding supports the presence of the multiple stellar populations in the bulge associated with the double RC. We note, however, that the absence of overall O-depletion of Na-rich stars remains questionable because the Na-O anticorrelation is generally observed in typical GCs. As we discussed, this could be due to the different metallicity domains between GCs ([Fe/H] $\lesssim$ -0.5 dex) and the bulge ([Fe/H] $\gtrsim$ -0.5 dex). Still more observations for O abundance and chemical evolution study for the bulge stars are necessary.

\begin{figure*}
\centering
\includegraphics[width=0.80\textwidth]{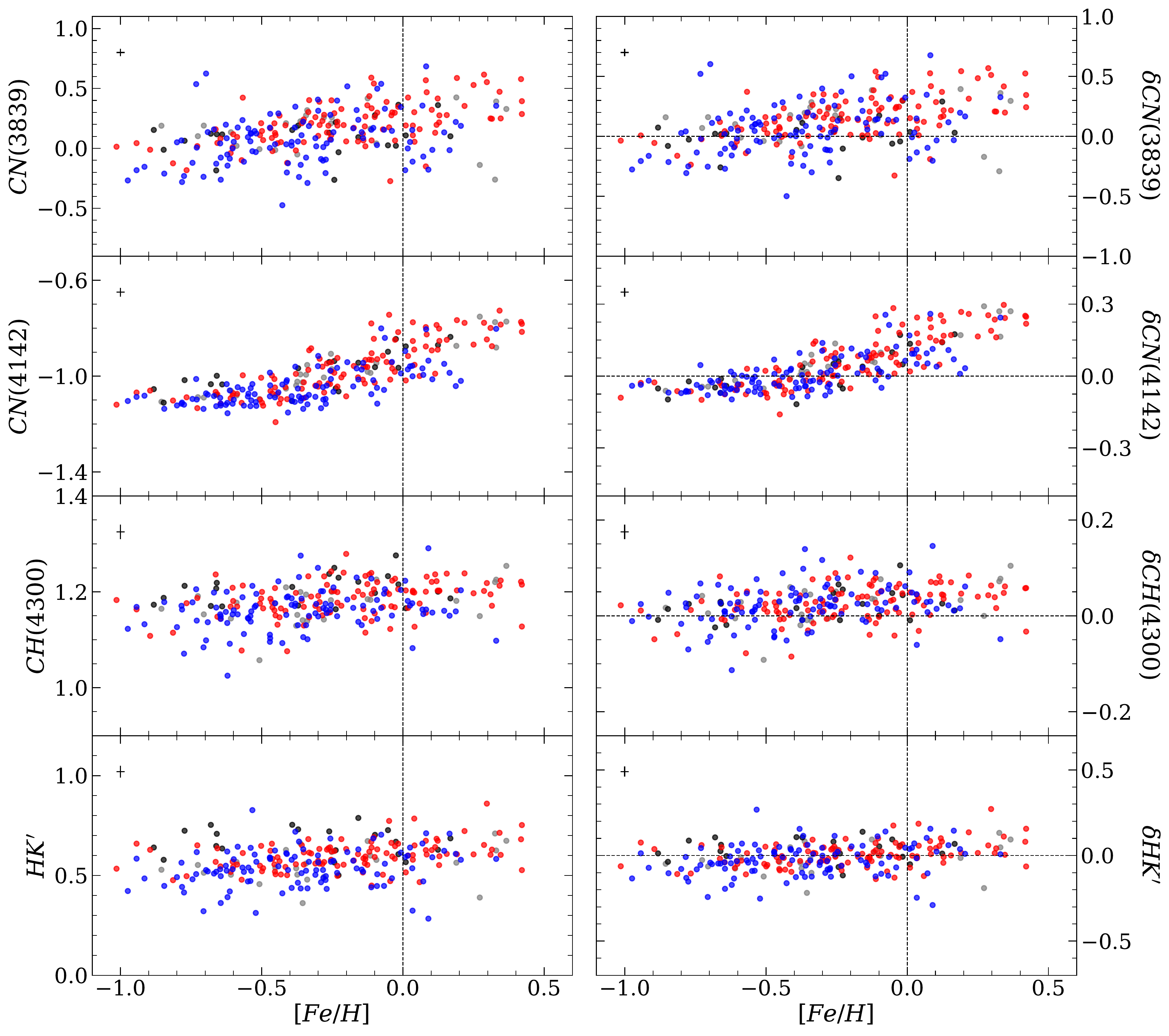}
\figcaption{
Spectral indices for commonly observed stars from low- and high-resolution spectroscopic studies as a function of [Fe/H]. Stars in the bRC group are more enhanced in the CN(3839), CN(4142), CH(4300), and HK$'$ indices compared to those in the fRC group (left panels) due to the effect of magnitude (see text). While these trends do not appear in $\delta$-index diagrams (right panels), both original and $\delta$-indices are correlated with [Fe/H]. The typical measurement error is plotted in the upper left corner of each panel. 
\label{fig:comp}
}
\end{figure*}


\section{Comparison with low-resolution spectroscopy} \label{sec:lrs}
Low-resolution spectroscopy to probe CN and CH molecular bands has long been used to study chemical properties of stellar populations in GCs \citep[e.g.,][]{Norris1987}. It is now well established that CN and CH bimodalities are one of the typical features of multiple stellar populations in GCs, together with the Na-O anticorrelation \citep{Kayser2008, Smolinski2011,Koch2019}. In particular, the CN and CH bands have the advantage that they can be easily measured from low-resolution spectra of faint stars. \citet{Lee2018} found a significant difference in the CN band strength between the stars in the bRC and fRC of the bulge and suggested the origin of the double RC in multiple populations. Here we compare our high-resolution spectroscopy with the low-resolution data of \citet{Lee2018} for 256 stars in common (107 on the bRC and 103 on the fRC). 

\begin{figure}
\centering
\includegraphics[width=0.45\textwidth]{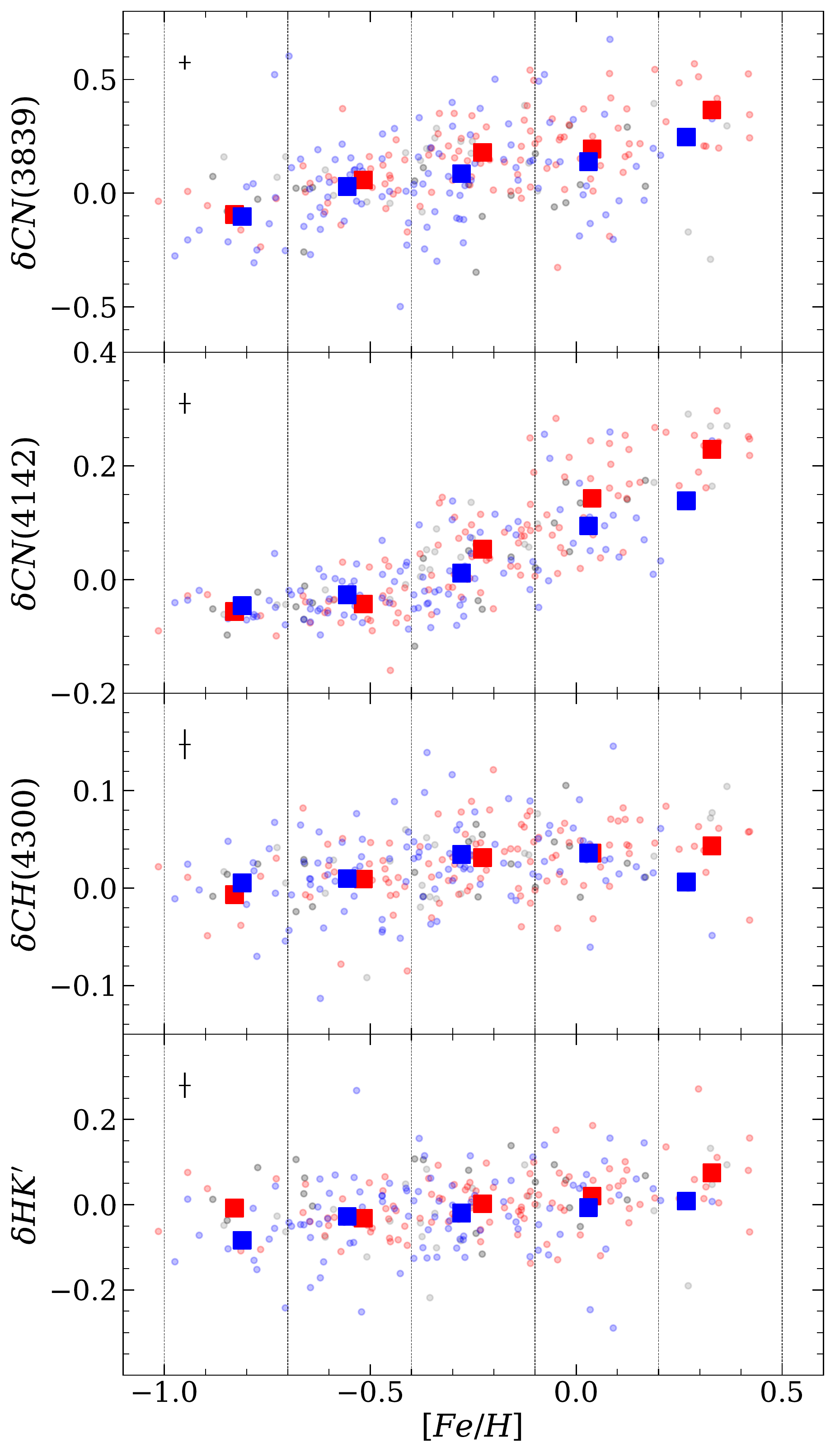}
\figcaption{
$\delta$CN(3839), $\delta$CN(4142), $\delta$CH(4300), and $\delta$HK$'$ indices (background circles) as a function of [Fe/H] with mean values for the bRC and fRC groups (red and blue squares) in each metallicity bin. Vertical dotted lines divides each range of metallicity at [Fe/H] = $-$1.0, $-$0.7, $-$0.4, $-$0.1, +0.2, and +0.5 dex. The bRC group generally has a higher mean value of $\delta$CN indices than the fRC group, while no significant differences are seen in $\delta$CH and $\delta$HK$'$ indices. 
\label{fig:CN}
}
\end{figure}

\begin{figure}
\centering
\includegraphics[width=0.45\textwidth]{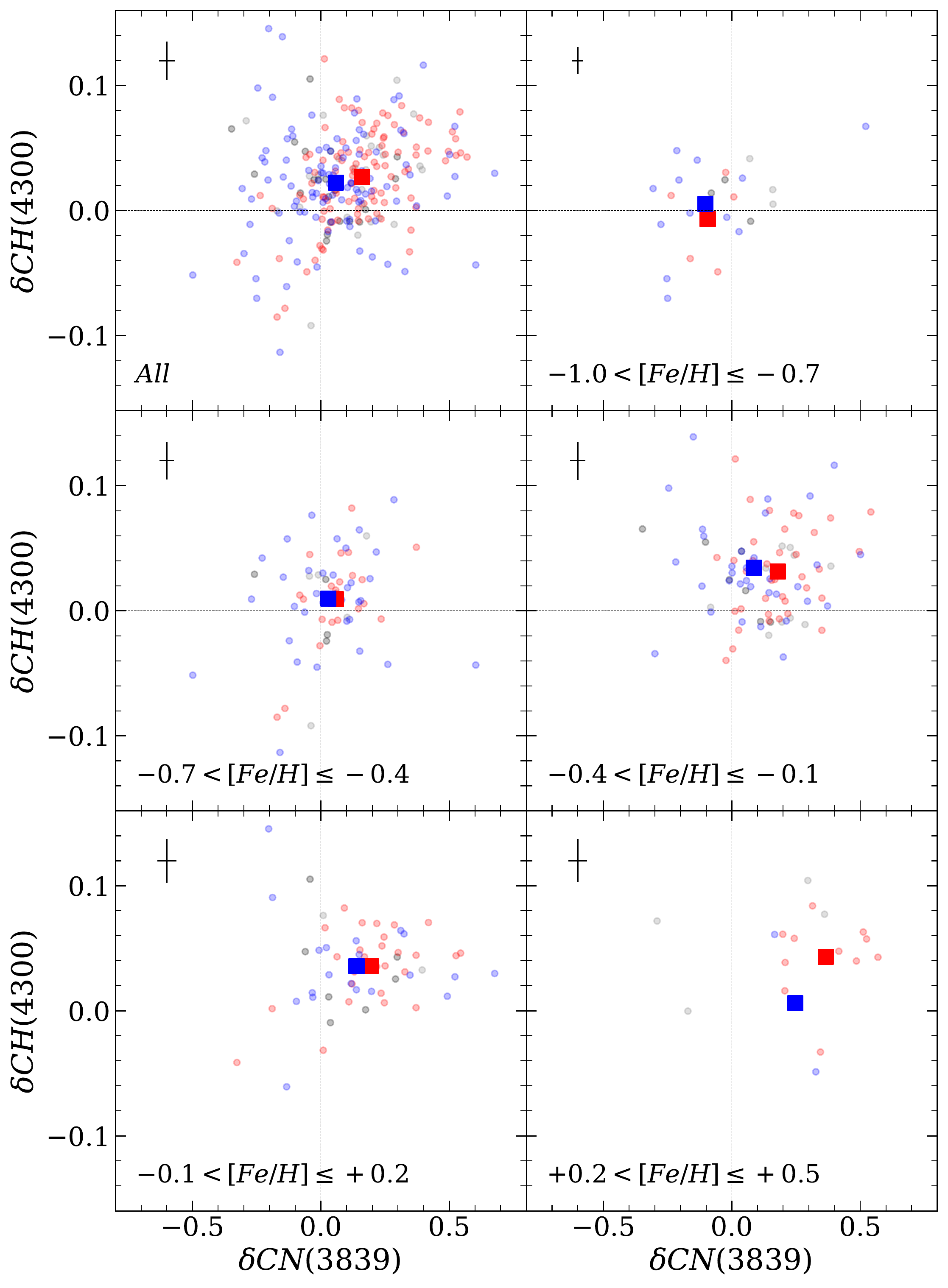}
\figcaption{
Distribution of stars on the $\delta$CN-$\delta$CH plane within different ranges of [Fe/H]. The symbols are the same as in Figure~\ref{fig:CN}. While both $\delta$CN and $\delta$CH indices of stars increase with increasing metallicity, the mean values of bRC and fRC groups show a flat-correlation in each [Fe/H] bin, where the bRC is enhanced in $\delta$CN, but not in $\delta$CH.
\label{fig:CN_CH}
}
\end{figure}

Figure~\ref{fig:comp} shows the spectral indices (CN(3839), CN(4142), CH(4300), and HK$'$) for bulge stars obtained from low-resolution spectroscopy as a function of [Fe/H] as measured from our high-resolution spectroscopy. The CN, CH, and HK$'$ indices are respectively related to N, C, and Ca abundances \citep[e.g.,][]{Smith1996}. These indices are increased, however, with decreasing magnitude, because they are also affected by temperature and surface gravity of stars. Therefore, stars in the bRC group are more enhanced in the CN(3839), CN(4142), CH(4300), and HK$'$ indices than those in the fRC (left panels of Figure~\ref{fig:comp}) due to this magnitude effect (see Figures~4 and 5 of \citealt{Lee2018}). The previous study therefore employed $\delta$-indices ($\delta$CN(3839), $\delta$CN(4142), $\delta$CH(4300) and $\delta$HK$'$), which are estimated as the difference between the original index and the least-squares fit of the index vs. magnitude, determined from RGB stars, to reduce this effect. The both original and $\delta$-indices, however, are also correlated with [Fe/H] (see Figure~\ref{fig:comp}). Thus, the difference in the CN band strength might be affected by the difference in [Fe/H] because the stars in the bRC group are more enhanced in [Fe/H] abundance than those in the fRC group (see Section~\ref{sec:metal}). In order to investigate this, we compare the mean values of the $\delta$-indices between the stars in the bRC and fRC groups at five different ranges of [Fe/H], each with 0.3 dex bin size. As shown in Figure~\ref{fig:CN}, the stars in the bRC group are generally more enhanced in the CN indices than the stars in the fRC, while there is no difference between the two groups for stars with [Fe/H] $\leq$ $-$0.7 dex. In the range of $-$0.4 $<$ [Fe/H] $\leq$ $-$0.1, where the most stars are located, the mean differences are estimated to be 0.093 mag in $\delta$CN(3839) index and 0.042 mag in $\delta$CN(4142) index, which are comparable to those obtained from the full sample of \citet[][$\Delta\delta$CN(3839) = 0.125 mag and $\Delta\delta$CN(4142) = 0.052 mag]{Lee2018}. Therefore, we can confirm that the stars in the two groups have indeed different chemical compositions of the CN band. In the case of the CH index, no statistically significant difference is observed between the two groups in all ranges of [Fe/H]. We note, however, that the CH-band could be saturated at high metallicity regime ([Fe/H] $>$ -0.7; see \citealt{Boberg2016}). Although stars in the bRC group are slightly more enhanced in the $\delta$HK$'$ index than the fRC group in some metallicity ranges ([Fe/H] $\leq$ $-$0.7 or [Fe/H] $>$ +0.2), it is hard to conclude due to the insufficient number of stars in each bin. 

As mentioned above, the correlation between the CN and CH indices can be a useful tool to study the origin of multiple stellar populations. While typical GCs show a CN-CH anticorrelation, GCs with heavy element variations, such as NGC~1851 and M22 \citep[e.g.,][]{Marino2015}, show a flat or positive CN-CH correlation that is mainly driven by SNe enrichment \citep{Lim2015, Lim2017}. In Figure~\ref{fig:CN_CH}, our sample stars are plotted in the CN-CH plane within each [Fe/H] bin. It is well demonstrated that the overall distribution of stars moves to the upper right according to increasing metallicity. However, the trend between the bRC and fRC is rather flat, where the bRC is enhanced in CN but not in  CH, similar to the case of NGC~1851 (see Figure~6 of \citealt{Lim2017}). This trend is still observed among stars in the limited range of [Fe/H], and most prominent in the range of $-$0.4 $<$ [Fe/H] $\leq$ $-$0.1 with a Pearson correlation coefficient of 0.070 and a $p$-value of 0.559. We note that a positive coefficient (0.0 $\sim$ 1.0) indicates positive correlations while a negative coefficient (-1.0 $\sim$ 0.0) indicates an anticorrelation. \citet{Lim2017} suggested that a metallicity variation with a flat or positive CN-CH correlation can be explained by a significant contribution of SNe enrichment to the formation of the metal-rich population. Therefore, similar chemical properties between the bRC in the bulge and the metal-rich population in the peculiar GCs, where  CN is enhanced but  CH is not, could be due to the same effect of SNe enrichment. This interpretation is also similar to the SNe enrichment to the formation of the metal-rich population in Terzan~5 \citep[][see also Section~\ref{sec:other}]{Origlia2011}. 


\section{Discussion} \label{sec:dis}
We have shown that stars in the bRC and fRC regimes in a high-latitude field towards the MW bulge ($l \sim -1 \degree$, $b \sim -8.5 \degree$) have significantly different chemical compositions, particularly in [Fe/H]. The stars belonging to the bRC group are more enhanced in [Fe/H] and [Na/Fe] but appear to be depleted in [Al/Fe] and [O/Fe] compared to those in the fRC group, although more observations are required to confirm the differences in [Na, Al, O/Fe] abundances. These chemical patterns between the bulge bRC and fRC populations are comparable to those observed between metal-rich and metal-poor populations in the peculiar bulge GC Terzan~5. We also find a number of Na-rich stars, which are most likely candidates for G3 stars. From the comparison with previous low-resolution spectroscopy, we confirmed the difference in CN-band strength with similar strength in CH band between the stars in the bRC and fRC groups regardless of [Fe/H]. Our results therefore appear to support the multiple population scenario for the origin of the double RC.

One may argue that the metallicity difference between the two RCs can be explained by an X-shaped bulge, as well by the vertical metallicity gradient observed in the inner MW \citep[e.g.,][]{Zoccali2008, Gonzalez2013}. Although this vertical metallicity gradient was originally thought to be a signature of the classical bulge, recent $N$-body simulations  suggest that a pseudo bulge model can also produce the vertical gradient \citep{Martinez-Valpuesta2013, Debattista2017}. In the X-shaped scenario for the double RC, where the bRC and fRC stars are  located in the foreground and background arms of the X-structure, respectively, the fRC would be further away from the Galactic plane compare to the bRC. Thus, the fRC stars should have lower metallicity than the bRC stars according to the vertical gradient. The effect of a vertical metallicity gradient, however, is far insufficient to reproduce the observed difference in [Fe/H] between the two RCs. For instance, a difference in vertical distance of $\sim$2 kpc is required to explain $\Delta$[Fe/H] $\sim$ 0.55 dex between the bright and faint RC stars, when applying a vertical gradient of 0.28 dex/kpc \citep{Gonzalez2013}, which is larger than predicted by present bulge models. 

On the other hand, two other possibilities for the difference in metallicity have to be considered. One is the contamination from the metal-rich Galactic bar component to the bRC regime. Although we expect a negligible contribution of the bar in this high-latitude field of the bulge, if some metal-rich bar stars are embedded only in the bRC regime, this might produce a metallicity difference between the two RCs. The other hypothesis is that the metal-rich component of the bulge is consisting the double RC, while the metal-poor component are mainly placed on the fRC regime. In this case, the mean metallicity of stars in the fRC would be lower than that of the bRC by dilution of metal-poor stars. An indication of this is already visible in Figure~17 of \citet{Johnson2020}. These two possibilities will be clarified by the further photometric or spectroscopic studies for the various fields of the bulge.

The difference in the metallicity between the bright and faint RCs is comparable to those observed in GCs showing heavy element variations. In particular, the chemical signatures of the double RC are closely analogous to Terzan~5, together with a similar feature on the CMD \citep{Ferraro2009}. It is reasonable in the respect that Terzan~5 is a metal-rich stellar system located in the bulge, while other peculiar GCs are relatively metal-poor ([Fe/H] $<$ -1.0). All the similarities between the double RC in the bulge and Terzan~5 suggest the association of their origin. One possible explanation is that a significant fraction of the outer bulge was formed from disrupted Terzan~5-like stellar systems in the early stage of the MW formation, and the bRC and fRC stars in the bulge would correspond to the the metal-rich and metal-poor populations originating in these systems. However, there remain several questions to be answered. As we discussed in Section~\ref{sec:metal}, Terzan~5 is a very peculiar stellar system, which has three distinct metallicity groups with different $\alpha$-elements abundances \citep[see, e.g.,][]{Origlia2011,Origlia2019,Massari2014}. Although the double RC of the bulge shows similar chemical trends of Fe, Na, Al, O, and $\alpha$-elements with Terzan~5 excepts for its most metal-poor population, a clear separation of metallicity between the two RCs is not observed. This could be explained if various metallicities of Terzan~5-like stellar systems are mixed in the bulge. Another stellar system analogous to Terzan~5, however, has not been discovered to date. Recently, \citet{Baumgardt2020} reported the absence of low mass GCs in the inner parts of the MW compared to the outer halo. They claimed that this could be either due to the short lifetime of low mass clusters owing to their sensitivity to the tidal field in the inner Galaxy or due to the difficulty of finding these clusters by large reddening and strong field star contamination \citep[see][]{Minniti2017,Palma2019}. In this regard, Terzan~5 could be one of the most massive survivors, while most of the other fragments were fully dissolved or not discovered yet. On the other hand, newly reported peculiar properties of some GCs, such as Gaia~1 and Liller~1\footnote{In the case of Gaia~1, three different mean metallicities are reported from three studies, respectively, with a large variation in [Fe/H] \citep[$\sim$ 0.6 dex;][]{Mucciarelli2017,Simpson2017,Koch2018}. Although there is a debate whether it is an intermediate-age GC or an old open cluster, the presence of three different metallicity populations is suspected in this cluster, like Terzan 5. In addition, Liller~1 is one of the most metal-rich and massive GCs in the bulge, which shows similar properties with Terzan~5 \citep[see, e.g.,][]{Saracino2015}.}, strengthen the possible existence of another Terzan~5-like stellar system.

In addition, the young age of the metal-rich population in Terzan~5 should be considered \citep[$\sim$4.5Gyr;][]{Ferraro2016}. As mentioned in Sections~\ref{sec:other} and \ref{sec:lrs}, it is likely that the metal-rich population in the bulge and Terzan~5 would have affected by SNe enrichment on a longer time scale. A detailed study of slow-neutron capture process ($s$-process) elements will provide useful information on the formation process of these systems. While there is no abundance measurement of $s$-process elements for stars in Terzan~5, it is well known that the metal-rich population of peculiar GCs is also enhanced in $s$-process elements \citep[see, e.g.,][]{Marino2011,Cordero2015,Marino2015}. Since the $s$-process enrichment is mainly attributed by AGB stars over a longer period of time \citep{Busso1999,Simmerer2004}, this could probe for an age gap between the formation of metal-poor and metal-rich populations. Thus, if a systematic difference in $s$-process elements is found in both Terzan~5 and the double RC of the bulge, it will indicate age difference among multiple populations and further strengthen the link between these two stellar systems. Although few La lines are located in our spectral coverage, we could not derive its abundances due to the low S/N ratio. Therefore, more high-quality spectroscopic data for the $s$-process elements will be helpful to clarify the formation mechanism of the double RC in Terzan~5 and the bulge.
 
In conclusion, while the MW bulge has both classical and pseudo bulge characteristics, it appears that the double RC feature is at least partially due to the different chemical compositions of the bright and faint RC stars on the multiple population scenario. Their chemical signatures indicate the possibility that a substantial fraction of the outer bulge would have formed by the assembly of Terzan~5-like stellar systems, in the form of a ``GC-origin'' or a  ``clump-origin'' bulge \citep[see][]{Lee2019, Inoue2012}. Our ongoing spectroscopy for RC and RGB stars in various fields of the bulge, as well as extensive photometric and spectroscopic surveys, will help to figure out the more detailed assembly history of the MW bulge.


\vspace{5mm}
We thank the referee for a number of helpful suggestions. We also thank Narae Hwang, Edward Olszewski, Anthony Kremin, Matthew Walker, and the LCO staff for observing support. Support for this work was provided by the National Research Foundation of Korea (grants 2017R1A6A3A11031025, 2017R1A2B3002919, and 2017R1A5A1070354). DL and AK acknowledge support from the Deutsche Forschungsgemeinschaft (DFG, German Research Foundation) -- Project-ID 138713538 -- SFB 881 (``The Milky Way System'', subprojects A03, A05, A11). DL thanks Sree Oh and Hyejeon Cho for comments and encouragements

\vspace{5mm}

\bibliographystyle{aasjournal}

\end{document}